%% file: frisch.tex
\documentclass[12pt,preprint]{aastex}
\usepackage{natbib}

\input{defcommand}

\slugcomment{Submitted to the Astrophysical Journal}

\shorttitle{LISM Velocity Distribution}
\shortauthors{Frisch, Grodnicki, Welty}

\begin{document}
\title{The Velocity Distribution of the Nearest Interstellar Gas}

\author{Priscilla C. Frisch}
\author{Lauren Grodnicki}
\author{Daniel E. Welty}
\affil{University of Chicago, Department of Astronomy and Astrophysics, 5460 S. Ellis Avenue, Chicago, IL 60637}

\begin{abstract}

The bulk flow velocity for the cluster of interstellar cloudlets
within $\sim$30 pc of the Sun is determined from optical and ultraviolet 
absorption line data, after omitting from the sample stars with 
circumstellar disks or variable emission lines and the active variable HR 1099.
Ninety-six velocity components towards the remaining 60 stars yield a streaming 
velocity through the local standard of rest of --17.0$\pm$4.6 \kms, with an 
upstream direction of \glong=2.3\deeg, \glat=--5.2\deeg\ (using
$Hipparcos$ values for the solar apex motion).
The velocity dispersion of the interstellar matter (ISM) within 30 pc is consistent with 
that of nearby diffuse clouds, but present statistics are inadequate
to distinguish between a Gaussian or exponential distribution about the bulk flow
velocity. 
The upstream direction of the bulk flow vector suggests an origin associated with the Loop I supernova remnant.
Groupings of component velocities by region are seen, indicating
regional departures from the bulk flow velocity or possibly separate clouds.
The absorption components from the cloudlet feeding ISM into the solar system 
form one of the regional features.  The nominal gradient between the velocities
of upstream and downstream gas may be an artifact of the Sun's location 
near the edge of the local cloud complex.
The Sun may emerge from the surrounding gas-patch within several thousand years.

\end{abstract}

\section{Introduction}

A number of studies have searched for correlations between the velocity of interstellar gas observed 
inside the solar system and towards external stars 
\citep[][]{AdamsFrisch:1977,McClintocketal:1978,LallementBertin:1992}.  
\citet[][]{AdamsFrisch:1977} showed that the velocity of interstellar gas 
inside the solar system differs by several \kms\ from interstellar cloud
velocities towards stars located in the upstream direction.
The discovery that the \HI\ \Lalpha\ line is redshifted by several \kms\ from the \DI\ line 
towards the nearest star ($\alpha$ Cen, 1.3 pc) confused
the identification of the cloud velocity in this direction \citep{Landsmanetal:1984}.
This shift has since been successfully modeled by including the
\Lalpha\ absorption from compressed \HI\ in the heliosheath \citep{LinskyWood:1996,Gayleyetal:1997}.
However, the heliosheath is observed only in the strong \Lalpha\ line.
The \DI, \MgII, and \FeII\ line velocities 
towards $\alpha$ Cen \citep[$\sim$--19 \kms,][]{Lallementetal:1995,LinskyWood:1996}
disagree by $\sim$2 \kms\ with interstellar gas
and dust velocities found inside the solar system 
\citep[e.g.][]{WellerMeier:1981,Witte:1996,Flynne:1998,Frischetal:1999}
when projected towards $\alpha$ Cen.  Optical and ultraviolet (UV) data show
that the bulk flow of the closest interstellar material (ISM) has an upstream
direction towards the Loop I supernova remnant, and
a velocity $\sim$20 \kms\ in the local standard of rest (LSR)
\citep{Frisch:1981,Crutcher:1982,Frisch:1995}.  This bulk flow is similar to 
ISM expanding around OB associations \citep[e.g.][]{Munch:1957}.
Following \citet[][hereafter SF02]{SlavinFrisch:2002}, the ISM within 30 pc is referred to
as the complex of local interstellar clouds (CLIC),
while the cloudlet feeding ISM into the solar system is denoted the Local Interstellar Cloud (LIC).
Optical \CaII\ and UV absorption data show that the CLIC is inhomogeneous on
subparsec scales, and that 
multiple absorption components are present towards the nearest stars
\citep[e.g.][]{MunchUnsold:1962,Ferlet:1986,LallementVidalMadjarFerlet:1986,Vallergaetal:1993,Lallementetal:1994,CrawfordDunkin:1995,WeltyCa:1996,Frisch:1995,Frisch:1996,CrawfordCraigWelsh:1997,CrawfordLallementWelsh:1998}.
The kinematics of the CLIC provides the opportunity to probe the history
of a diffuse cloud, the relevance of small scale structure to ISM physics,
and to gauge the past and future galactic environment of the Sun \citep{Frisch:1997}.

Nearby ISM provides a unique set of constraints for determining diffuse cloud physics.
Observations of both pickup ions inside the solar system
(formed by interactions of interstellar neutrals with the solar wind)
and interstellar absorption lines towards nearby stars 
have been used to constrain the first full radiative transfer model of nearby ISM
(SF02).
These results indicate that the interstellar properties at the solar location are
T$\sim$7,000 K, \nHI$\sim$0.24 \cc, \ne$\sim$0.13 \cc, 
and fractional ionizations $X$(H)$\sim$31\% and $X$(He)$\sim$48\%,
with both density and ionization levels varying towards the cloud surface 
(Model 17).
The model includes emission from a conductive interface, which yields
an excess of helium compared to hydrogen ionization (SF02).
If the density \nHI$\sim$0.24 \cc\ is typical for all cloudlets near the
Sun, then \NHI\ towards $\alpha$ Cen \citep[1.3 pc,][]{LinskyWood:1996} suggests a 
filling factor $f\sim$0.4.
Enhanced refractory element abundances in warm nearby gas provide evidence that 
local ISM has been shocked \citep{Frischetal:1999}.
The presence of cold \CaII\ and \NaI\ absorption components (Doppler 
width $b_{\rm D}\leq$0.8 \kms) towards $\alpha$ Pav (56 pc) and 
$\delta$ Cyg (52 pc) indicate at least 
an order of magnitude variation for ISM temperatures within 50 pc (see
references in Table \ref{tab-comps}).

The sensitivity of heliosphere properties \citep{ZankFrisch:1999} ---
and astrospheres in general \citep[including extra-solar
planetary systems,][]{Frisch:1993a} --- to the physical conditions of the 
surrounding ISM motivates this exploration of small
scale structure in the nearby ISM using cloud velocity as a structure proxy.
Understanding small scale structure in the CLIC
will also help decipher the signature that astrospheres leave on the
interstellar \HI\ Ly$\alpha$ absorption line.
This paper focuses on the bulk flow kinematics of ISM within $\sim$30 pc of the Sun.
Anticipating a conclusion, the CLIC appears to consist of
an ensemble of cloudlets at velocities consistent with a 
random distribution about a mean bulk flow velocity.
The first results of this study were presented by \citet{Frisch:2001b}.

\section{Velocities of Absorption Components \label{streaming}}

\subsection{Method}

In this analysis we assume that the motions of nearby
interstellar clouds (i.e. the CLIC) in the local standard of rest (LSR) can be described by a linear flow vector, \Vflow,
which is characterized by the flow velocity and the galactic coordinates
of the direction
from which the gas flows (implying that the flow velocity is $<$0).
Our approach is to calculate a best fitting flow vector for a set of
observed interstellar cloud component radial velocities which sample interstellar 
gas within $\sim$30 pc (Section \ref{sbulk}), and then to determine the 
distribution of component velocities about this bulk flow vector.  
If the CLIC (d$<$30 pc) is part of an
expanding shell feature from Loop I (Section \ref{loopI}), it
will subtend a total angle equivalent to about 10\% of the expanding
shell and yield $<$5\% deviations from a linear flow velocity (i.e. $\sim$1 \kms\ for
uniform expansion).  Therefore, over the scale length of the CLIC, the
linear assumption is sufficient.  The bulk flow velocity vector is determined from a
fitting procedure that varies \Vflow\ in order to minimize the sum,
$\Phi_{\rm m}$, over m observed interstellar absorption line
components, of the square of the difference between the projected flow
and observed velocity towards each star, i.e., to minimize:
\begin{equation}
\Phi_{\rm m } = \sum_{\rm i=1}^{\rm m} ~ \dVi^2 , \\
\end{equation}
where
\begin{equation}
\dV_{\rm i } = \Viobs - \Vflow \cdot \hat{k}_{\rm star}  . \\
\label{eqn2}
\end{equation}
For a flow vector calculated from  a set of m components (\Vflow {\rm (m)}),
\begin{equation}
\dV_{\rm i} {\rm (m)} = \Viobs - \Vflow {\rm (m)} \cdot \hat{k}_{\rm star} ,  \\
\label{eqn3}
\end{equation}
where \dVi(m) is the scalar radial velocity (at the solar location) of component i in the rest 
frame of \Vflow(m).
Here $ \hat{k}_{\rm star}$ is the unit position vector towards each
star and \Viobs\ is the observed heliocentric (HC) radial velocity
for an interstellar velocity component towards that star (here denoted i,
where i=1,m).  The fits were performed using the FindMin function in Mathematica, which
determined the local minima of the function $\Phi_{\rm m}$ as a
function of the three variables V,l,b which define \Vflow.  Note that
since many stars show more than one interstellar absorption component,
the number of components (m) is greater than the number of stars.  An
unweighted fit was employed since there is no obvious weighting
function.  Line broadening is not suitable since instrument
resolutions and thermal broadening vary (artificially blending
velocity components).  Column densities are unsuitable since they vary between
species.  Note that since we are interested in the distribution of the
deviation of component velocities in the rest frame of \Vflow, the component data set
includes all measured components for each star.  We find that this
procedure yields a predicted flow velocity, \Vflow, which is
relatively insensitive to the detailed set of the selected velocity components,
provided that sky coverage is good and that several stars with
components at anomalous velocities are removed from the sample (next section).

In principle, the full three-dimensional flow vector \Vflow\ is
calculated correctly for cases where the perturbations from the flow
(\dVi) are random, since the Sun is immersed in the flow.  
Alternative cases whereby perturbations in the flow have a directional
dependence (or preference), or where the flow is decelerated \citep{Frisch:1995}, 
are not considered here.  However, we make an elementary attempt to
identify systematic regional patterns in the velocity components.

\subsection{Component Set and Bulk Velocity \label{sbulk}}

The velocities of optical and ultraviolet absorption lines towards
nearby stars, and towards more distant stars which sample primarily
the CLIC gas, are used to evaluate the velocity distribution of the
CLIC.  Interstellar absorption line data for 67 stars which sample the
CLIC are listed in Table \ref{tab-ref}, with component information
in Table \ref{tab-comps}.  This component set forms the ``unrestricted sample''.  
A range of data sources was used in order to cover as much of the sky as possible.
Observations of the optical interstellar \CaII\ lines represent the 
primary source of cloud velocity data, but the optical data do not adequately sample the
downstream region where column densities are low over the 
first $\sim$30 pc \citep[e.g. $N$(\CaII)$<< 10^{10}$ \cmtwo,][]{BruhweilerKondo:1982,FrischYork:1983,Frisch:1995,Lallementetal:1995}. 
Observations of the intrinsically stronger UV lines of \FeII, \MgII,
\DI, and \HI\ are available for several stars in the downwind
direction and towards the north galactic pole (where column densities
are also low).

The component sample results primarily from fits to optical and UV absorption
lines observed at relatively high resolution ($<$3 \kms); 
medium resolution UV data were used for a few stars.  When both optical
and UV data exist, the optical \CaII\ data (resolution 0.5--3 \kms)
are the first choice, and the UV data second choice since about 50\% of
the optical spectra were acquired with resolution FWHM$<$1.5 \kms.  
Component velocities are generally obtained from a process which iteratively
fits each discernible component in the absorption line profile with a population of atoms with a Maxwellian
velocity distribution about the central component velocity, with a
projection in the radial line-of-sight towards the Sun of \Vobs\
(denoted \Viobs\ in eqs. \ref{eqn2}, \ref{eqn3}).  The number of
components and the component descriptors (column density, \Vobs, and
Doppler broadening constant, $b_{\rm D} $) are sensitive to instrument resolution and
signal-to-noise and to the judgment of the observer.  

The component sample used here is not of uniform quality.
For example, the star $\alpha$ Oph (14 pc) has been observed at
resolutions of 0.3 \kms\ by \citet{CrawfordDunkin:1995} and 1.2 \kms\
\citet{WeltyCa:1996}.  The 0.3 \kms\ resolution profiles were fitted with
four components with velocities --32.0$\pm$0.5, --28.4$\pm$0.3,
--26.2$\pm$0.2 and --23.6$\pm$0.5 \kms.  The 1.2 \kms\ resolution
profiles were fitted with three components with velocities --33.03 \kms,
--26.25 \kms, --22.16 \kms.  While temporal variations in interstellar \CaII\
absorption profiles are seen in a few cases \citep[e.g.][]{Hobbsetal:1991}, 
comparisons between the two absorption profiles suggest that
the component at --28.4 \kms\ was unresolved in the lower resolution data.

The derived flow vector depends on the component
set used in the fitting procedure, and the unrestricted sample is small enough
that lines incorrectly identified as interstellar may 
alter the results.  Therefore, 
the basic star sample was restricted to omit stars 
with either identified circumstellar disks or strong emission line spectra
(where misidentification of circumstellar features is possible).
Thus the stars $\epsilon$ Eri (3.2 pc), $\beta$ Pic (19.3 pc), $\beta$ Leo (11.0 pc), $\alpha$ Lyr
(7.6 pc), $\alpha$ PsA (7.7 pc), which have circumstellar disks identified at 60 $\mu$m \citep[e.g.][]{Habingetal:2001}, 
were omitted from the sample.
Although interstellar components can in principle be distinguished
from circumstellar disk components since stellar velocities are
known, we considered the sample to be more reliable when stars with circumstellar
disks are omitted.  The emission line star $\alpha$ Eri (44 pc) was also omitted,
as it shows variable \HeI\ and \MgII chromospheric 
features indicating that weak circumstellar absorption features may be present.
The active RS CVn variable HR 1099 \citep{Woodetal:1996}
was also omitted from the sample because the inclusion of the components
towards this star yielded results that are inconsistent with
fits towards the other stars (see below).
With the above omissions, the remaining restricted sample consists
of 60 stars with a total of 96 components.  Star distances range from
1.3 pc ($\alpha$ Cen) to 132 pc ($\epsilon$ CMa).  
Components towards the two stars beyond 70 pc, $\epsilon$ CMa and 31 Com, 
were included because these stars primarily sample ISM within 30 pc
\citep{GryJenkins:2001,Dringetal:1997} and fill a gap in the spatial coverage.

This basic component sample was fitted for the flow vector \Vflow,
yielding a HC vector \Vflow(96) with radial velocity --28.1$\pm$4.6 \kms,
flowing from the upstream direction \glong=12.4\deeg, \glat=11.6\deeg.
The uncertainty is the 1-$\sigma$ value of the velocity 
component distribution about the bulk flow velocity.
Figure \ref{fig1}a shows the deviations \dVi(96) plotted
against the stellar distance (omitting the two stars
more distant than 70 pc).  
Note that 76 of these 96 components originate in the upstream direction,
while the remaining components originate downstream, reflecting the
higher column densities of nearby ISM found upstream \citep[e.g.,][]{Frisch:1995}.

The positions of the stars forming the 96 component set are plotted in Figure \ref{fig2}a.
Most of the sky is well sampled, except for the interval
\glong=150\deeg--180 \deeg\ in the direction of the third galactic quadrant void
(also known as the Puppis window or the $\beta$ CMa tunnel), 
where mean interstellar column densities are low
out to $\sim$100 pc ($\overline{n_{\rm \HI }}<$0.003 \cc).
Figure \ref{fig3}a shows \dVi(96)\ as a function of the galactic longitude 
of the star; no strong systematic spatial dependence is seen.  
For comparison, the LSR velocities of the 96 components (\dVi(LSR)) 
are plotted against galactic longitude in Figure \ref{fig3}b, 
illustrating the expected behavior for an incorrect vector describing the flow.
Throughout this paper we use the solar apex motion based on $Hipparcos$
data, corresponding to a solar motion towards the galactic 
coordinates \glong=27.7\deeg, \glat=32.4\deeg\ at the velocity \V=13.4 \kms\
\citep{DehnenBinney:1998}.  

The robustness of \Vflow(96) was tested in several ways.  The first test was
to restrict the star sample to stars within 29 pc
of the Sun (except for $\epsilon$ CMa and 31 Com, which sample the nearest
ISM and are required for completeness in low column density
directions).  This restriction gives 46 components towards 31 stars.  
The best fit HC flow vector for this 46 component sample, \Vflow(46), 
corresponds to an upstream direction of \glong=12.5\deeg, \glat=12.5\deeg, 
with an inflow velocity of --28.1$\pm$4.3 \kms.  The fact that \Vflow(46) 
is virtually identical to \Vflow(96) is not a coincidence, since the initial
selection of stars in Table \ref{tab-comps} was restricted to stars
which did not appear to contain neutral ISM more distant than $\sim$30
pc.  The closeness of the \Vflow(96) and \Vflow(46) vectors support the validity
of eliminating HR 1099 from the component sample, since had HR 1099 been retained 
in the restricted sample the resulting fitted vector (\Vflow(99)) would have
differed from \Vflow(46) by 0.6 \kms\ and 2\deeg.  

The second test was a search for spatially correlated variations in \dVi(96).
Star positions are replotted in Figure \ref{fig2}b, with symbols coded
for \dVi(96).
No systematic positional dependence appears for components with $| \dVi(96) | >$5 \kms.
However, when the components with $| \dVi(96) |  <$5 \kms\ are plotted with
symbols which code the
sign of \dVi(96), stars in the interval \glong=150\deeg $\rightarrow$250\deeg\
are found to show components with systematically small but negative \dVi(96)
values (Figure \ref{fig2}c, note several components are overplotted).
The best-fitting flow vector, \Vflow(20), for
the 20 components toward 14 stars located in the interval \glong=150\deeg $\rightarrow$250\deeg\ corresponds to a vector velocity --25.8$\pm$4.3 \kms\
from the upstream direction \glong=6.2\deeg, \glat=10.4\deeg.  If only a single
component per star is selected for the fit, biasing towards components
with small \dVi(He) values, the resulting fitted vector \Vflow(14)
corresponds to a velocity --25.6$\pm$1.3 \kms\ from the upstream
direction \glong=6.4\deeg, \glat=13.1\deeg, which is
close to \Vflow(He) (Table \ref{tab-vec}, Section \ref{lic}).
The dispersion of the same 14 components around \Vflow(He) is $\pm$1.7 \kms.
Gas near the LIC velocity (Table \ref{tab-vec}) dominates in the downwind 
direction (l$\sim$180\deeg), as found previously.  
The LIC and other regional variations in \Vflow(96) are discussed in Section
\ref{regions}.

\subsection{Component Velocity Dispersion \label{dispersion}}

The randomness of interstellar cloud velocities has long been of considerable
interest, with early arguments presented for random motions of
$\sim$20 \kms\ for interstellar particles  \citep{Spitzer:1942},
compared to recent data showing root-mean-square velocity dispersions 
of $\sim$0.5 \kms\ in individual cold clouds.
Early studies of diffuse cloud kinematics found an exponential distribution
for cloud velocities, suggestive of a turbulent ISM \citep{Blaauw:1952,Munch:1957}.

The functional form of the velocity distribution of the CLIC gas was 
tested, but the results prove inconclusive.  A plot of the number of 
components (N, ordinate) binned for a given value of \dVi(96) (abscissa) is
shown in Figure \ref{fig4}.  Here, N is determined with \dVi(96) 
binned with 1 \kms\ bin sizes.  The histogram is ``noisy'' because 
the sample is small. 

For a purely random distribution of individual components about the central flow velocity \Vflow(96),
the form of the \dVi\ distribution should be Gaussian:
\begin{equation}
\Psi ( V )=  \frac{c_{\rm g}}{b \surd \pi }{\rm exp}-(V-V_o)^2/b^2
\label{eq-gaus}
\end{equation}
The \citet{Blaauw:1952} and \citet{Munch:1957} results for
\CaII\ and \NaI\ lines, based on low resolution data (FWHM$>$5 \kms),
are consistent with an exponential distribution for cloud velocities:  
\begin{equation}
\Psi ( V )=  \frac{c_{\rm e}}{2\eta}{\rm exp}-|V-V_o|/\eta  ,
\label{eq-psi}
\end{equation}
such as is expected for a turbulent flow.  Here, $b=\sigma \sqrt 2 $, 
$c_{\rm g}$ and $c_{\rm e}$ are normalizing constants, 
and $\sigma$ and $\eta \sqrt 2$ are the root-mean-square deviations
of the Gaussian and exponential distributions respectively.

The 96 component sample was tested for the distribution of \dVi(96), 
assuming alternatively Gaussian and exponential distributions.
The best-fit Gaussian distribution yields \glat=6.2 \kms,
and the best-fit exponential distribution yields
$\eta$=5.1 \kms.  The present data do not distinguish
between these two distributions (see overplotted functions in Figure \ref{fig4}).
Blaauw concluded that an exponential function with mean speed 
$\eta \sim$5$\pm$1 \kms\ provided the best fit
to \CaII\ K-line data towards 300 stars observed by \citet{Adams:1949} with
instrumental resolution $\sim$9 \kms.
Munch inferred that an exponential form fit observations of the \CaII\ and \NaI\
doublets towards 112 stars, with $\eta  \sim$5--6 \kms\ for \CaII\ in the
Orion arm and a somewhat smaller value for \NaI.
Munch concluded that turbulence dominates the velocity distribution of these
components.  
These early low resolution data undersample blended
components at low velocity and therefore probably enhance component statistics
in the distribution wings with respect to the central region. 
It is therefore puzzling that we find a similar value from higher
resolution data.  This issue is discussed further in a subsequent paper
\citep{FrischWelty:2002}.

\section{Discussion \label{discussion}}

\subsection{Upstream Direction and Loop I\label{loopI}}

The best-fit heliocentric velocity vector, \Vflow(96), corresponds to a bulk velocity
\mbox{--28.1} \kms\ flowing from the upstream direction \glong=12.4\deeg, 
\glat=11.6\deeg\ (Section \ref{sbulk}).
However in order to compare  \Vflow(96) with morphological features in the 
astronomical sky the solar apex motion through the LSR must be subtracted.
Subtracting the solar apex motion from \Vflow(96) gives a
``true'' upstream direction (in LSR) in galactic coordinates
of \glong=2.3\deeg, \glat=--5.2\deeg, with flow velocity --17.0 \kms.\footnote{For comparison, the standard solar apex
motion (based on the brightest stars with a range of ages) is 19.7
\kms\ towards \glong=57\deeg, \glat=+22\deeg, which yields an upstream LSR
flow motion of --19.4 \kms\ from the direction \glong=331.4\deeg,
\glat=--4.9\deeg.}
The \Vflow(96) upwind direction is shown plotted against the filamentary
\HI\ structure associated with Loop I \citep[21 cm data, Figure \ref{fig-sxrb},][]{Hartmann:1997}.  
The position of the radio continuum Loop I
shell, defined from the discovery 408 MHz radio continuum data, is 
marked \citep[from][]{Berkhuijsen:1971,Haslam:1982}.

Low column densities in the CLIC components (\NH$< 10^{18.5}$ \cmtwo, typically)
prevent linking CLIC velocities to individual features in the 21-cm \HI\ sky maps.
The relation between the flow of interstellar gas past the Sun and the
Loop I superbubble has been discussed from several viewpoints 
\citep[e.g.][]{Frisch:1979,Frisch:1981,Crutcher:1982,Frisch:1995,Crawford:1991,deGeus:1992}.
The Loop I superbubble remnant (the brightest segment of which is the 
North Polar Spur) is centered near \glong$\sim$320\deeg,
\glat$\sim$5\deeg, with distance $\sim$130 pc and radius
$\sim $80\deeg\ \citep[based on \HI\ 21 cm data,][]{Heiles:1998}.  
The radio-continuum Loop I is
centered at \glong=329\deeg, \glat=+17.5\deeg, radius 58\deeg, and is strongly 
limb-brightened in the tangential direction near \glong$ \sim$35\deeg\ \citep{Berkhuijsen:1973}.
The center of Loop I defined from \HI\ data is offset from the radio-continuum
center by $\sim$15\deeg.
The \HI\ shell shows a polar cap at \Vlsr$\sim$--30 \kms\ and
\glong,\glat$\approx$300\deeg,--10\deeg\ \citep{Heiles:1989}, where
the cold component seen at --19.6 \kms\ towards $\alpha$ Pav may originate (Table \ref{tab-comps}).
This expansion velocity implies a projected velocity 
of $\approx$--14 \kms\ for the \Vflow(96) LSR upwind direction, versus the 
best-fit LSR velocity of --17.0 \kms.
The filamentary structure and likely asymmetric expansion of the \HI\ shell 
make it difficult to identify the approaching
portion of the \HI\ shell, so the $\sim$20\% difference between the projected
polar cap velocity and the LSR bulk velocity of the CLIC gas possibly is within uncertainties.
However, the \HI\ shell regions traced by the 21-cm
polar cap emission are denser and colder than the CLIC gas appears to be.  
Alternatively, models of shell expansion from the formation of the Sco-Cen 
Association Upper Scorpius subgroup
place an \HI\ shell with LSR velocity $\sim$--22 \kms\ and age $\sim$4 Myrs at the solar location
\citep{Crawford:1991,deGeus:1992, Frisch:1995}.\footnote{Note the age given for the most recent shell-forming event should be 4 Myrs, not 400,000 yrs \citep[][page 532]{Frisch:1995}.}

\subsection{Comparison with Previous Models \label{previous}}

The bulk flow of nearby interstellar gas
has been determined from optical absorption lines observed towards nearby stars 
\citep[primarily \CaII, e.g.][]{Crutcher:1982,Frisch:1986,LallementVidalMadjarFerlet:1986,LallementBertin:1992,Vallergaetal:1993,Frisch:1995,Frisch:1997}.
The velocity vectors determined from these earlier studies are summarized in Table \ref{tab-rev} and shown in Figure 7.
Since column densities in the downstream direction are low (\logNHI$<$18 \cmtwo), 
optical lines do not easily sample this interval.  This causes bulk flow
vectors based primarily on \CaII\ data to be strongly weighted towards
upstream gas and  quite sensitive to the details of the star sample.
The inclusion of UV data provides an adequate sample of downstream gas, but at lower resolution (3 \kms).
\citet[][LB92]{LallementBertin:1992} used UV data to identify
the ISM in the downstream direction, denoting it the 
``anti-Galactic'' cloud (AG, Table \ref{tab-rev}), which is the same as the LIC.
ISM identified in the upstream direction, generally towards the galactic-center
hemisphere, was denoted the ``Galactic'' cloud (G, Table \ref{tab-rev}).
The velocity difference between the G and AG clouds is smaller than found
by Adams and Frisch (1977), based on backscattered \Lalpha\ \HI\ data,
because of the deceleration and compression of \HI\ in the heliosheath 
regions (which was unknown in 1977).
LB92 concluded that either the Sun is located in a
patch of gas with velocity intermediate between the G and AG clouds, or that \HI\
is decelerated in the heliosheath (or both).  
The $Ulysses$ He and 584 \AA\ backscattering data confirm the
velocity of ISM inside the solar system, so that the
general velocity difference between the upstream and downstream ISM is real.
However, the fact that
the velocities of CLIC cloudlets are consistent
with a random distribution about the mean bulk flow velocity
suggests that the upstream/downstream velocity gradient locally displayed by the G versus LIC 
vectors may be an artifact of the location of the Sun with respect to the
``edge'' of the CLIC complex.  Alternatively, the presence of a blue-shifted cloud towards CMa
\citep{Lallementetal:1994,GryJenkins:2001} is also consistent with a velocity
gradient.

For comparison with the LB92 G-AG two-flow model, a separate bulk flow velocity
was calculated for downstream components.  The 20 LIC components originating 
in the interval \glong=150\deeg $\rightarrow$250\deeg\ were removed from the 96 component sample.
The fit to the resulting 76 components yields \Vflow(76)=--29.3$\pm$4.0 \kms, from HC upstream direction \glong=13.1\deeg, \glat=11.2\deeg.
The velocities of \Vflow(76) and the G vector (\Vflow(G)) coincide, although the
upstream directions differ by $\sim$12\deeg.
The observed 3.5 \kms\ difference between \Vflow(76) and \Vflow(20) and
the near coincidence of upwind directions (within $\sim$8\deeg)
suggests a deceleration of the bulk flow.  The two-flow model with the two vectors \Vflow(20) and \Vflow(76)
(or alternatively with \Vflow(He) and \Vflow(G))
is not a unique description of CLIC gas.
For example, the velocity dispersion of \Vflow(76) is about equal to the difference between
the \Vflow(76) and \Vflow(20) velocities ($\sim$4 \kms).  
The second shortcoming is that
6 stars in the \glong=150\deeg $\rightarrow$250\deeg\ interval show two
absorption components, and the second components are not accommodated by
the two-flow model.  The dispersion of the entire 20 components about \Vflow(96) is
5.4 \kms.  The dispersion around \Vflow(He) of the 14 components best matching the
LIC velocity (1.5 \kms) is only slightly better than the dispersion of these
components about \Vflow(96) (1.6 \kms).

\subsection{Regional Properties\label{regions}}

The small filling factor of nearby ISM, combined with the component distribution,
led to attempts to understand the positions of nearby `cloudlets'
\citep[e.g.][]{LallementVidalMadjarFerlet:1986,Frisch:1996}, the clouds 
towards $\alpha$ Oph and $\alpha$ Aql \citep{MunchUnsold:1962,Frisch:1981,Ferlet:1986}, and the ``shape'' of the cloud
around the solar system \citep{Frisch:1996,RedfieldLinsky:2000}.
These earlier approaches used velocities grouped either in the observed
heliocentric velocity, or in the LIC velocity.  
Here we use an alternative approach, and evaluate cloud membership in the rest frame of \Vflow(96).  
Using the \dVi(96) values, we identify regional clumps of ISM based on 
similar \dVi(96) values for stars sampling a relatively compact region of the sky, 
except for the LIC which is identified by \dVi(He) (or \dVi(20)).
The component groups that appear to represent subsets of the flow
are listed in Table \ref{tab-cld} and are discussed in order below.
The positions of the clouds are plotted in Figure 6 as symbols overlying
star positions.  

\subsubsection{1.  Cloud Surrounding the Solar System \label{lic}}

The velocity of the interstellar cloud surrounding the solar system, the LIC,
has been determined from observations of the resonance fluorescence of the 
584 A transition from interstellar \HeI\ inside the solar system
\citep[e.g.][]{WellerMeier:1981,Flynne:1998} and from $Ulysses$ $in ~situ$ 
measurements of interstellar \HeI\ 
\citep{Witte:1996}.  Since the trajectories of interstellar H atoms
inside the solar system are subject to radiation pressure, these \HeI\ data
yield the best LIC velocity.
The $Ulysses$ $GAS$-detector data give \Vflow(He)=\mbox{--25.3$\pm$0.5}
\kms, \glong=3.9$\pm$1.0\deeg, \glat=15.8$\pm$1.3\deeg\ (and
$T$=6,550$\pm$1,050 K) for the HC upstream direction 
\citep[][and private communication]{Witte:1996}, which is within the
uncertainties of the \HeI\ 584 \AA\ backscattered radiation vector \citep{Flynne:1998}.
When this He motion is converted to a vector in the local standard of rest (LSR)
using the $Hipparcos$ solar apex motion \citep{DehnenBinney:1998},
the upstream direction corresponds to a velocity of
--14.7 \kms, arriving from the upwind direction \glong=345.9\deeg, \glat=--1.1\deeg.
In the LSR, this differs by $\sim$2 \kms\
and $\sim$20\deeg\ in direction from the bulk velocity vector.
Pure deceleration of the flow would not alter the upstream direction, so shear
motions may be present in the bulk flow.

$Copernicus$, $IUE$ and $Hubble Space Telescope$ ($HST$) observations of interstellar \DI, \FeII, \MgII, 
towards $\alpha$ Cen (1.3 pc) indicate a HC velocity for the ISM
in the range \mbox{--17.7$\pm$0.1} to --18.2$\pm$0.1 \kms,
versus the value --16.1 \kms\ predicted by the projection of the 
$Ulysses$ vector \citep{Landsmanetal:1984,Lallementetal:1995,LinskyWood:1996}.
This difference has been interpreted as indicating that the cloud surrounding
the solar system terminates within $\sim$10,000 au of the Sun in the
direction of $\alpha$ Cen. 
The $\alpha$ Cen cloud shows \dVi(He)\mbox{$\sim$--2.8 \kms} (and \dVi(96)=--3.8 \kms).
This evidence for a cloud boundary close to the Sun in the direction
of $\alpha$ Cen is consistent with column densities measured towards the 
nearest stars, which indicate that $<$40\% of space is filled with ISM for
volume density \nHI$\sim$0.24 \cc\ as found for the LIC (SL02).

Thirty CLIC components have velocities consistent with the velocity of interstellar
He inside the solar system (\Vflow(He)) using the criterion $|$\dVi(He)$| <$1.5 \kms.  
However, components towards stars beyond 15 pc (Figure \ref{fig1}b)
have predominantly negative \dVi(He) values since these stars
are generally located in the upstream direction.
The stars showing these 30 \Vflow(He) components
are distributed relatively uniformly across the sky (Figure \ref{fig2}d).
Refitting the 30 components essentially reproduces
\Vflow(He), and does not alter the distance dependence, suggesting the more
distant components with $|$\dVi(He)$|  <$1.5 \kms\ either are blends of local and
distant gas or are formed in independent cloudlets with no local contribution.
The five stars within 15 pc with \Vflow(He) components are $\alpha$ CMa (2.7 pc), 
61 CygA (3.5 pc), $\epsilon$ Ind (3.6 pc), 40 EriA (5 pc), and $\alpha$ Aur (13 pc). 
The UV absorption lines towards $\alpha$ CMi (3.5 pc) have been fitted with
both single-component and two-component models \citep{Linskyetal:1995}.  
With the single-component fit (Table \ref{tab-comps}), \dVi(He)=1.8 \kms;
however \dVi(20)=0.5 \kms\ so that \Vflow(20) provides a better fit for local gas towards this star.
(A second component redshifted by $\sim$2.6 \kms\ (with $\sim$50\%
of the column density of the main component) is not coincident with either flow vector.)

The velocity vector of the cloud within the solar system provides poor
agreement with the restricted sample of 96 components, as seen in
Figure \ref{fig3}c where \dVi(He) is plotted for all 96 components.
The fact that $|$\Vflow(He)$| < |$\Vflow(96)$|$ yields the effect that
nearby ISM in the downstream direction 
is blueshifted in the rest frame of \Vflow(96).

Several studies have noted the small distance to the upstream cloud
edge for the ISM surrounding the solar system ($< 10^4$ au, e.g., LB92).
The relative Sun-cloud velocity ($\sim$25 pc/10$^6$ yrs)
suggests that the galactic environment of the Sun will vary within the 
next $\sim$2,000 years \citep{Frisch:1997,WoodLinskyZank:2000}.

The LIC is reliably identified only in the 
\glong=150\deeg$\rightarrow$250\deeg\ interval where slightly more distant
gas is absent, and in this interval \Vflow(He) is an excellent descriptor of 
component velocities.
The inability to identify the LIC reliably in sidestream and upstream 
directions indicates the two-flow model must be evaluated with caution 
in most sightlines, and in particular where velocity information is unavailable.
In addition, since the CLIC represents a cluster of comoving clouds with
a velocity dispersion of $\sim$5 \kms, 
simple closed surface LIC models which can be topologically deformed to a sphere
may not realistically represent the LIC cloudlet.  

\subsubsection{Other Regional Component Groups}

\paragraph{2.  Blue Cloud:  \label{bc}} 
The ``Blue Cloud'' (BC) has been identified towards $\alpha$ CMa and $\epsilon$ CMa,
placing the cloud within 3 pc of the Sun
\citep[][GJ]{Lallementetal:1994,Hebrardetal:1999,GryJenkins:2001}. 
The BC has \dVi(96)$\sim$--10 \kms, and is blueshifted from the LIC by $\sim$6.5 \kms. 
The filling factor of the LIC towards
$\alpha$ CMa is $f \sim 0.3$, suggesting that the BC is a spatially separate feature.
Towards $\alpha$ CMa the BC appears to be cooler and denser than the 
LIC \citep{Hebrardetal:1999}, while towards $\epsilon$ CMa it appears warmer and denser than the LIC (GJ).

\paragraph{3.  Aquila-Ophiuchus Cloud: \label{ao}}
The Aquila-Ophiuchus cloud (AOC) is disclosed by a set of \CaII\ components with \dVi(96)=--6.1$\pm$0.9 \kms\ towards stars 
located in the interval \glong=28\deeg\ to 48\deeg, \glat=--5\deeg\ to +23\deeg.
The stars $\alpha$ Aql, $\alpha$ Oph, $\zeta$ Aql, $\gamma$ Oph, and $\lambda$ Aql
show the Aql-Oph cloud component, and the 5.1 pc distance of $\alpha$ Aql places the cloud close to the Sun.
For each of these five background stars, the component formed in the AOC is
the weakest and most blue-shifted component seen towards the star.
The similarity between the AOC velocity and the velocity of weakest and most 
blueshifted component towards $\delta$ Cyg (\dVi(96)=--6.9 \kms) may be a 
coincidence since the the $\delta$ Cyg component originates in a cold cloud 
\citep[\bdoppler=0.47 \kms, 0.42 \kms\ for \CaII, \NaI\ respectively,][]{WeltyCa:1996}, 
and cold clouds are not expected within 5 pc.
The AOC is located near the solar apex direction, \glong=27.7\deeg, 
\glat=32.4\deeg.
If the unobserved tangential velocity component is $\sim$0 \kms,
the Sun may encounter this cloud within the next $\sim$160,000 years.
Components from this cloud are seen clearly in Figure 3a.

\paragraph{4.  Pegasus-Aquarius Cloud: \label{pa}}
The Pegasus-Aquarius cloud, with \dVi(96)=4.2$\pm$0.5 \kms, is seen 
towards stars grouped in the interval  \glong=75\deeg $\pm$13\deeg,
\glat=--44\deeg $\pm$5\deeg.  Background stars are $\theta$ Peg, $\alpha$ Peg, $\gamma$ Aqr, $\eta$ Aqr, and this cloud must be within 30 pc of the Sun.

\paragraph{5.  North Pole Cloud: \label{upstream}}
A group of components with velocity 1.7$\pm$0.6 \kms\ is found towards stars 
at high galactic latitudes, \glat$>$53\deeg, including the Ursa Majoris region.
The nearest star with a component in this group is $\alpha$ CrB, at 23 pc.

\paragraph{6.  South Pole Cloud: \label{sp}}
A cloud with \Vflow(96)=2.4$\pm$1.0 \kms\ occupies an irregularly
shaped region covering the 
South Galactic Pole (\glat$\leq$48\deeg).  The South Pole Cloud (SPC) cloud
is seen towards $\epsilon$ Ind, $\alpha$ Hyi, $\tau ^3$ Eri, $\beta$ Cet.
However, two of these components (towards $\epsilon$ Ind and $\beta$ Cet)
also have $|$\dVi(He)$|  <$0.1 \kms, indicating they also could be formed in 
the LIC (which is seen towards two other low latitude stars, 40 Eri and EP Eri).
If so, the SPC is not a separate cloudlet.
The nearest star in this sample is $\epsilon$ Ind at 3.6 pc, indicating 
the SPC, if real, is quite nearby.  Additional data are required to
test for the SPC.

\paragraph{7.  Filamentary-like Feature: \label{ring}}

Perhaps the most puzzling of the component groups suggests a 
filamentary structure, with \dVi(96)= +6.9$\pm$1.5 \kms.
Since this feature has a large angular extent ($\sim$90\deeg, Figure \ref{fig6}),
it represents a single cloudlet only if the tangential (unobserved) velocity 
is $\sim$0 \kms\ for all of the stars, similar to a fragment of an 
expanding ring.  This feature is at \glong=284\deeg $\pm$12\deeg, \glat=--4\deeg $\pm$50\deeg, and 
it is seen towards $\delta$ CrV, $\gamma$ CrV,  HR 4023, $\delta$ Vel, $\alpha$ Hyi.

\section{Conclusions}

The principal conclusions of this paper are that:

1.  The bulk flow velocity for interstellar matter within $\sim$30 pc of the Sun
is determined from optical and UV absorption line data.  A self-consistent 
flow vector is determined if stars with infrared-emitting circumstellar disks
($\epsilon$ Eri, $\beta$ Pic, $\beta$ Leo, $\alpha$ Lyr, and $\alpha$ PsA),
variable emission lines ($\alpha$ Eri), and HR 1099 are omitted from the sample.
Fits to the remaining 96 component sample (60 stars)
yield a bulk flow velocity (\Vflow(96)) through the LSR (using $Hipparcos$
data on the solar apex motion) of magnitude --17.0 \kms, with an upstream 
direction of \glong=2.3\deeg, \glat=--5.2\deeg\ (Table \ref{tab-vec}).  
In the heliocentric velocity frame (i.e. with respect to the Sun),
$|$\Vflow(96)$|$=28.1 \kms, approaching the Sun from the upstream 
direction \glong=12.4\deeg, \glat=+11.6\deeg\ (the 1-$\sigma$ velocity 
uncertainty is $\pm$4.6 \kms, based on the
component distribution about the bulk flow velocity).

2.  The CLIC gas (Cluster of Local Interstellar Clouds) is defined by this ensemble of velocity
components, which show a Gaussian or exponential distribution in the rest
frame of the central flow velocity, and the dispersion of this distribution 
is typical for dispersions determined from lower resolution observations of 
diffuse clouds.  The dispersion in the velocities of the individual CLIC 
components partaking in the bulk flow explains the difference between
the velocity of the interstellar cloud inside the solar system
(25.3 \kms, HC) versus the overall bulk flow velocity (--28.1 \kms, HC).

3.  This LSR upstream direction suggests the CLIC gas may be part of a
superbubble shell expanding from the Loop I supernova remnant,
or from an earlier epoch of superbubble formation in the Scorpius-Centaurus Association.
The cold absorption component formed in the polar cap of Loop I appears to be present in front
of $\alpha$ Pav.

4.  The velocity of the interstellar cloud observed inside the solar system,
and in the downstream direction towards nearby stars (\Vflow(He)) differs 
both from \Vflow(96), and from the velocity of ISM towards
the nearest star $\alpha$ Cen.
However, 85\% of the observed components consistent with \Vflow(He)
are probably formed in unrelated parts of the CLIC.  This complexity suggests that
simple smooth closed-surface models for local interstellar gas are unlikely 
to be accurate in detail.  The components within 1.5 \kms\ of the velocity of interstellar He
within the solar system (\Vflow(He)) dominate the galactic longitude
interval \glong=150\deeg $\rightarrow$250\deeg, suggesting a deceleration 
of the flow in the third galactic quadrant.  The Sun is likely to emerge
from the gas-patch now surrounding it within the next several thousand years.

5.  Several spatially distinct groups of components sharing a common \dVi(96)
suggest the bulk flow is composed of cloudlets (Table \ref{tab-cld}),
including the previously known LIC and Blue Cloud in the downstream
direction.   Distinct cloudlets in the upstream direction include the
Aquila-Ophiuchus cloud with components at 
at \dVi(96)$\sim$--6 \kms.  This cloud may extend to within
5 pc of the Sun since it is seen towards $\alpha$ Aql.
Components in this cloud form the weakest and highest velocity \CaII\
component for each star in which it is observed.
This cloud is in the solar apex direction, and 
if the non-radial component of the cloud velocity is small
the Sun will encounter this cloud in $<$200,000 years.
A distinct cloud is seen towards Pegasus/Aquarius, at \dVi(96)=3.8$\pm$0.3 \kms.
Another distinct cloud is found within 22 pc towards the
North Galactic Pole, with \dVi(96)=1.7$\pm$0.6 \kms.

6.  These results show the kinematics of CLIC components are consistent with 
early studies showing macroscopic turbulence in the ISM 
within $\sim$500 pc \citep{Blaauw:1952,Munch:1957}.  Low filling factors 
for nearby ISM($<$40\%), and mean cloudlet speed $\eta \sim$5 \kms\ 
in the flow reference frame (\Vflow(96)) suggest that the CLIC is a cluster of comoving cloudlets.
If this velocity dispersion is due to turbulence (as the larger
samples considered by Blaauw and Munch suggest), then it is distinct from the
microscopic turbulence which broadens absorption lines towards nearby stars
($b ^2$=$2kT/m$ + 2$V_{\rm turb} ^2$, where $V_{\rm turb} \sim$1 \kms). 

7.  A two-flow model of the velocity components is not a unique description 
of CLIC gas.  For example, the velocity dispersion of \Vflow(76) is about 
equal to the difference between the \Vflow(76) and \Vflow(20) velocities 
($\sim$4 \kms).  In addition, multiple components in the downstream gas are 
not accommodated by a two-flow model.  The result is that morphological models 
of the LIC must include velocity information for accurate results.

Studies of the CLIC, where individual cloudlets can be studied
in high resolution UV data, and in some sightlines in optical data, 
offer the best opportunity for reliably determining the relation between
the kinematical and spatial characteristics of diffuse interstellar clouds.

\acknowledgements
This research has been supported by NASA grants NAG5-6405 (Frisch), NAG5-8163 (Frisch), and NAG5-3228 (Welty).


\input{frisch.bbl}
\clearpage
\input{tab1}
\clearpage
\input{tab2}
\clearpage
\input{tab3}

\clearpage
\input{tab4}
\clearpage
\input{tab5}

\input{figurecaptions}
\end{document}

%% file: defcommand.tex
\def\bdoppler{\hbox{$b_{\rm Dop}$}}

\def\2S12{\hbox{$^{2}{\rm S }_{1/2}$}}

\def\V{\hbox{$V$}}

\def\Vlsr{\hbox{$v_{\rm lsr}$}}

\def\Lalpha{\hbox{Lyman${\alpha}$}}

\def\kms{\hbox{km s$^{\rm -1}$}}

\def\glong{\hbox{$l^{\rm II}$}}
\def\glat{\hbox{$b^{\rm II}$}}

\def\logNHI{\hbox{log[$N({\rm H^o})$]}}

\def\NH{\hbox{$N$(H)}}

\def\NHI{\hbox{$N$(\ion{H}{1}$)$}}

\def\nHI{\hbox{$n({\rm H^o})$}}

\def\cmtwo{\hbox{cm$^{-2}$}}

\def\cc{\hbox{cm$^{-3}$}}

\def\deeg{\hbox{$^{\rm o}$}}

\def\ne{\hbox{$n {\rm (e^- )}$}}
\def\HI{\hbox{${\rm H^o}$}}
\def\NHI{\hbox{$N {\rm (H^o )}$}}
\def\CaII{\hbox{Ca$^+$}}

\def\NaI{\hbox{Na$^{\rm o }$}}
\def\HeI{\hbox{He$^{\rm o }$}}

\def\DI{\hbox{D$^0$}}
\def\FeII{\hbox{Fe$^+$}}

\def\TiII{\hbox{Ti$^+$}}
\def\MgII{\hbox{Mg$^+$}}
\def\Lalpha{\hbox{L$\alpha$}}
\def\kms{\hbox{km s$^{-1}$}}
\def\cmtwo{\hbox{cm$^{-2}$}}
\def\dV{\hbox{d$V_{\rm  }$}}
\def\dVi{\hbox{d$V_{\rm i}$}}
\def\V{$V$}
\def\Vflow{\hbox{$\vec{V}_{\rm flow}$}}

\def\22V{\hbox{d$V^{\rm 101}_{\rm s}$}}

\def\Vobs{\hbox{$v_{\rm obs}$}}
\def\Viobs{\hbox{$v_{\rm i,obs}$}}

\def\bdoppler{\hbox{$b$}}

%% file: tab1.tex
\begin{deluxetable}{llll llll l}
\tablecolumns{8} 
\tablewidth{0pt} 
\tabletypesize{\small}
\tablecaption{Streaming Motions in HC and LSR Rest Frames\label{tab-vec}}
\tablehead{ 
\colhead{}&\multicolumn{3}{c}{HC Vector\tablenotemark{a}}&\colhead{}&\multicolumn{3}{c}{LSR Vector\tablenotemark{{\rm a,b}}}& \colhead{Sample\tablenotemark{c}} \\
\cline{2-4} \cline{6-8} \\ 
\colhead{}&\colhead{Vel.}&\colhead{l}&\colhead{b}&\colhead{}& \colhead{Vel} & \colhead{l}&\colhead{b}&\colhead{} \\
\colhead{}&\colhead{\kms}&\colhead{Deg.}&\colhead{Deg.}&\colhead{}&\colhead{\kms}&\colhead{Deg.}&\colhead{Deg.} & \colhead{} \\
}
\startdata 
\hline 
\sidehead{Bulk Fits:}
\Vflow(96)&--28.1$\pm$4.6&12.4&11.6&$~~~$&--17.0&2.3&--5.2&60 stars \\
\Vflow(46)&--28.1$\pm$4.3&12.5&12.5&$~~~$&&&& 31 stars, d$<$29 pc \\

\sidehead{Upstream:}
\Vflow(76)&--29.3$\pm$4.0&13.1&11.2&$~~~$&--18.1&3.9&--4.7&46 stars\\
\Vflow(G)&--29.4&4.5&20.5&$~~~$&&&&G-cloud \citep[][]{LallementBertin:1992}\\

\sidehead{Downstream, \glong=150$\rightarrow$250:}
\Vflow(He)&--25.3$\pm$0.5&3.9$\pm$1.0&15.8$\pm$1.3&&--14.7&345.9&--1.1&$Ulysses$ \HeI\ value \\
&&&&&&&& \citep[][]{Witte:1996} \\
\Vflow(20)&--25.8$\pm$4.3&6.2&10.4&$~~~$&&&&14 stars \\
\Vflow(584\AA)&--26.4$\pm$1.5&4.5$\pm$0.6&14.2$\pm$0.6&&&&&Backscattered \HeI\ 584 \AA\ \\
&&&&&&&&\citep{Flynne:1998} \\

\hline 
\enddata 
\tablenotetext{a}{Galactic coordinates correspond to upstream directions.
The velocity uncertainty is the 1-$\sigma$ value of the 
component distribution about the bulk flow velocity.}
\tablenotetext{b}{Based on Hipparcos solar apex motion \citep{DehnenBinney:1998}.}
\tablenotetext{c}{Sample stars are listed in Tables \ref{tab-ref}
and \ref{tab-comps}.}

\end{deluxetable} 

%% file: tab2.tex
\begin{deluxetable}{lll}
\tablecolumns{3} 
\tablewidth{0pc} 
\tabletypesize{\small}
\tablecaption{Data Sources\label{tab-ref}}
\tablehead{ 
\colhead{Reference}&\colhead{Nominal}&\colhead{Star List}\\
\colhead{}&\colhead{Resl.\tablenotemark{a}}&\colhead{}\\
\colhead{}&\colhead{\kms}&\colhead{}\\
}
\startdata 
\hline 
\sidehead{Optical Data:}
\citet[][FW02]{FrischWelty:2002}&1.3--1.5 &$\alpha$ CMa, $\alpha$ Leo, $\eta$ UMa, $\gamma$ UMa, $\delta$ UMa, $\beta$ Lib \\*
				&&$\alpha$ CrB, $\lambda$ Oph, $\delta$ Her, $\nu$ Ser, $\beta$ Ser, $o$ Ser, $\lambda$ Aql \\*
				&& $\zeta$ Aql, $\alpha$ Aql, $\beta$ Aur, $\alpha$ Peg \\
\citet[][CLW98]{CrawfordLallementWelsh:1998}&0.35&$\alpha$ And, $\delta$ Vel, $\alpha$ Pav, $\theta$ Peg \\
\citet[][CCW97]{CrawfordCraigWelsh:1997}&0.35&$\alpha$ Hyi, $\iota$ Cen, $\epsilon$ Gru, $\gamma$ Aqr, $\gamma$ Oph \\
\citet[][W96]{WeltyCa:1996}&1.2&  $\gamma$ CrV, $\delta$ CrV,  $\delta$ Cyg \\
\citet[][CD95]{CrawfordDunkin:1995}&0.3& $\alpha$ Oph, $\alpha$ Gru, $\alpha$ Eri \\
\citet[][V93]{Vallergaetal:1993}&1.9&$\alpha$2 CVn, $\eta$ Aqr, $\kappa$ And \\
\citet[][B93]{Bertinetal:1993}&3.0& $\tau$3 Eri, HR 4023\\
\citet[][LB92]{LallementBertin:1992}&3.0& $\alpha$ Cep, $\delta$ Cas, $\alpha$ Lac\\

\sidehead{Ultraviolet Data:}
\citet[][GJ01]{GryJenkins:2001}&2.7& $\epsilon$ CMa \\
\citet[][WLZ00]{WoodLinskyZank:2000}&2.7& 36 Oph \\
\citet[][W00]{Woodetal:2000}&$\sim$15& EP Eri, DX Leo, LQ Hya, V368 Cep \\
\citet[][Sa99]{Sahuetal:1999}&2.7&G191-B2B \\
\citet[][H99]{Hebrardetal:1999}&2.7&$\alpha$ CMa \\
\citet[][WL98]{WoodLinsky:1998}&3.5&61 Cyg, 40 Eri A \\
\citet[][D97]{Dringetal:1997}&3.5&$\epsilon$ Eri,  $\beta$ Gem,  $\alpha$ Tri,  $\sigma$ Gem,  31 Com \\
\citet[][WAL96]{WoodAlexanderLinsky:1996} &3.5& $\epsilon$ Ind, $\lambda$ And \\
\citet[][P97]{Piskunovetal:1997}&3.5& $\beta$ Cet \\
\citet[][Fer95]{Ferletetal:1995}&3.5& $\alpha$ PsA \\
\citet[][Lin95]{Linskyetal:1995}&3.5& $\alpha$ CMi, $\alpha$ Aur \\
\citet[][Lal95]{Lallementetal:1995}&3.5& $\alpha$ Lyr, $\beta$ Leo, $\beta$ Pic \\
\hline 
\enddata 

\end{deluxetable} 

%% file: tab3.tex
\begin{deluxetable}{llrrllllll}
\tablecolumns{10} 
\tablewidth{0pc} 
\tabletypesize{\small}
\tablecaption{Velocity Components of Unrestricted Sample \label{tab-comps}}
\tablehead{ 
\colhead{HD}& \colhead{Name}& \colhead{l}& \colhead{b}& \colhead{d}& \colhead{Spec.}& \colhead{N(X)\tablenotemark{a}}& \colhead{Velocity}& \colhead{\dVi(96)} & \colhead{Ref.\tablenotemark{b}}  \\
\colhead{}& \colhead{}& \colhead{deg.}& \colhead{deg.}& \colhead{pc}& \colhead{}& \colhead{}& \colhead{\kms}& \colhead{\kms}& \colhead{}  \\
}
\startdata 
   128621  & $\alpha$   CenB  &  315.7  &   --0.7  &  1.4  &   Glpl  &  3.89e17  &  --18.9 	   & 	  --3.8 	&   P97 	  \\ 
  48915  & $\alpha$  CMa  &  227.2  &   --8.9  &  2.6  &  A1V  &   2.5e17  &   13.7 	   & 	  --9.5 	&   H99 	  \\ 
    &      &    &    &    &    &  0.16e10  &   19.6 	   & 	  --3.6 	&  FW02 	  \\ 
 22049\tablenotemark{c,d}  & $\epsilon$  Eri  &   196.0  &   --48.0  &  3.2  &  K2V  &  7.50e17  &   21.3 	   & 	  --1.3 	&   D97 	  \\ 
  61421  & $\alpha$  CMi  &  213.7  &   13.0  &  3.5  &   F5IV/V  &  7.59e17  &   20.8 	   & 	  --3.0 	& Lin95 	  \\ 
   201092  &   61   CygA  &  82.3  &  --5.8  &  3.5  &  K5V  &  7.08e17  &   --9.0 	   & 	   0.0 	&  WL98 	  \\ 
    &      &    &    &    &    &  7.41e17  &   --3.0 	   & 	   6.0 	&  WL98 	  \\ 
   209100  & $\epsilon$  Ind  &  336.2  &  --48.0  &  3.6  &  K4.5V  &  1.00e18  &   --9.2 	   & 	   1.5 	& WAL96 	  \\ 
  26965  &   40   EriA  &   200.8 &   --38.0 &  5.0  &  K1V  &  8.71e17  &   21.6 	   & 	  --3.4 	&  WL98 	  \\ 
   155886  &   36  OphAB  &   358.3   &   6.9  &   6.0 &  K1V/K0V  &  7.08e17  &  --28.4 	   & 	  --1.2 	&  WLZ00 	  \\ 
   187642  & $\alpha$  Aql  &   47.7  &   --8.9  &  5.1  &  A7V  &   0.3e10  &  --26.9 	   & 	  --5.5 	&  FW02 	  \\ 
    &      &    &    &    &    &   0.5e10  &  --22.8 	   & 	  --1.4 	&  FW02 	  \\ 
    &      &    &    &    &    &   0.9e10  &  --18.1 	   & 	   3.3 	&  FW02 	  \\ 
 172167\tablenotemark{c,d}  & $\alpha$  Lyr  &   67.5  &   19.2  &  7.6  &  A0V  &   6.5e12  &  --18.3 	   & 	  --1.5 	& Lal95 	  \\ 
    &      &    &    &    &    &   6.9e12  &  --16.0 	   & 	   0.8 	& Lal95 	  \\ 
    &      &    &    &    &    &  1.03e13  &  --12.7 	   & 	   4.1 	& Lal95 	  \\ 
 216956\tablenotemark{c,d}  & $\alpha$  PsA  &   20.5  &  --64.9  &  7.7  &  A3V  &   3.8e12  &   --3.2 	   & 	   3.3 	& Fer95 	  \\ 
    &      &    &    &    &    &   1.8e12  &   --9.7 	   & 	  --3.2 	& Fer95 	  \\ 
  17925  &   EP  Eri  &   192.1   &   --58.3   &   10.4  &  K2V  &   8.9e17  &  19. 	   & 	  --0.3 	&   W00 	  \\ 
  62509  &  $\beta$  Gem  &  192.2  &   23.4  &   10.6  &   K0II  &  1.15e18  &   21.9 	   & 	  --1.2 	&   D97 	  \\ 
    &      &    &    &    &    &  6.82e17  &   33.0 	   & 	   9.9 	&   D97 	  \\ 
 102647\tablenotemark{c,d}  &  $\beta$  Leo  &  250.6  &   70.8  &   11.0  &  A3V  &  \nodata  &   --0.8 	   & 	  --0.3 	& Lal95 	  \\ 
    &      &    &    &    &    &  \nodata  &  2.7 	   & 	   3.2 	& Lal95 	  \\ 
    &      &    &    &    &    &  \nodata  &   13.3 	   & 	  13.8 	& Lal95 	  \\ 
  34029  & $\alpha$  Aur  &  162.6  &  4.6  &   12.9  & G5IIIepl  &  1.74e18  &   22.0 	   & 	  --1.4 	& Lin95 	  \\ 
   159561  & $\alpha$  Oph  &   35.9  &   22.6  &   14.3  &  A5III  &  1.00e11  &  --23.6 	   & 	   1.9 	&  CD95 	  \\ 
    &      &    &    &    &    &  1.58e11  &  --26.2 	   & 	  --0.7 	&  CD95 	  \\ 
    &      &    &    &    &    &  1.59e10  &  --32.0 	   & 	  --6.5 	&  CD95 	  \\ 
    &      &    &    &    &    &  6.31e10  &  --28.4 	   & 	  --2.9 	&  CD95 	  \\ 
   203280  & $\alpha$  Cep  &  101.0  &  9.2  &   15.0  &   A7IV  &   0.7e10  &  0.2 	   & 	   1.8 	&  LB92 	  \\ 
  432  &  $\beta$  Cas  &  117.5  &   --3.3  &   16.7  &   F2IV  &  1.51e18  &   10 	   & 	   2.5 	&   P97 	  \\ 
  11443  & $\alpha$  Tri  &  138.6  &  --31.4  &   17.5  &   F6IV  &  1.15e18  &   17.6 	   & 	   0.8 	&   D97 	  \\ 
    &      &    &    &    &    &  9.79e17  &   12.7 	   & 	  --4.1 	&   D97 	  \\ 
  82443  &   DX  Leo  &   201.2   &  46.1   &   17.7  &  K2V  &  5.e17  &  11. 	   & 	  --3.8 	&   W00 	  \\ 
   115892  &  $\iota$  Cen  &  309.4  &   25.8  &   18.0  &  A2V  &   2.5e10  &  --18.2 	   & 	  --4.5 	& CCW97 	  \\ 
  82558  &   LQ  Hya  &   244.6   &  28.4   &   18.3  &  K2V  &  5.62e18  &   6. 	   & 	  --6.2 	&   W00 	  \\ 
 39060\tablenotemark{c,d}  &  $\beta$  Pic  &  258.4  &  --30.6  &   19.3  &  A5V  &  \nodata  &   --3.2 	   & 	   --15.7 	& Fer95 	  \\ 
    &      &    &    &    &    &  \nodata  &   --9.7 	   & 	   --22.2 	& Fer95 	  \\ 
   220140  &    V368Cep  &   118.5   &  16.9   &   19.7  &  K2V  &   8.9e17  &   5. 	   & 	  --0.6 	&   W00 	  \\ 
  12311  & $\alpha$  Hyi  &  289.5  &  --53.8  &   21.9  &  F0V  &  2.88e10  &  9.8 	   & 	   7.3 	& CCW97 	  \\ 
     &      &    &    &    &    &  5.01e10  &  4.9 	   & 	   2.4 	& CCW97 	  \\ 
   139006  & $\alpha$  CrB  &   41.9  &   53.8  &   22.9  &  A0V  &  1.03e10  &  --17.4 	   & 	   1.3 	&  FW02 	  \\ 
  87901  & $\alpha$  Leo  &  226.4  &   48.9  &   23.8  &  B7V  &   0.5e10  &   10.5 	   & 	  --0.3 	&  FW02 	  \\ 
   156164  & $\delta$  Her  &   46.8  &   31.4  &   24.1  &   A3IV  &   3.1e10  &  --19.7 	   & 	   2.7 	&  FW02 	  \\ 
  74956  & $\delta$  Vel  &  272.1  &   --7.4  &   24.4  &  A1V  &  0.86e10  &   15.6 	   & 	  10.0 	& CLW98 	  \\ 
    &      &    &    &    &    &  2.38e10  &  1.3 	   & 	  --4.3 	& CLW98 	  \\ 
    &      &    &    &    &    &  4.34e10  &   11.6 	   & 	   6.0 	& CLW98 	  \\ 
   106591  & $\delta$  UMa  &  132.6  &   59.4  &   25.0  &  A3V  &   0.9e10  &  3.8 	   & 	   1.6 	&  FW02 	  \\ 
  40183  &  $\beta$  Aur  &  167.5  &   10.4  &   25.2  &   A2IVpl  &   1.0e10  &   22.3 	   & 	  --1.3 	&  FW02 	  \\ 
   177724  &  $\zeta$  Aql  &   46.9  &  3.3  &   25.5  &   A0Vn  &   0.6e10  &  --24.5 	   & 	  --1.5 	&  FW02 	  \\ 
    &      &    &    &    &    &   0.9e10  &  --20.6 	   & 	   2.4 	&  FW02 	  \\ 
    &      &    &    &    &    &   1.4e10  &  --29.7 	   & 	  --6.7 	&  FW02 	  \\ 
   103287  & $\gamma$  UMa  &  140.7  &   61.4  &   25.6  &  A0V  &   0.8e10  &  4.4 	   & 	   1.2 	&  FW02 	  \\ 
   222107  &  $\lambda$  And  &  109.9  &  --14.5  &   25.8  &  G8III  &  2.81e18  &  6.5 	   & 	   1.6 	& WAL96 	  \\ 
  18978  &  $\tau$3  Eri  &  213.5  &  --60.3  &   26.4  &  A4V  &   2.0e10  &   15.9 	   & 	  --1.7 	&   B93 	  \\ 
    &      &    &    &    &    &   3.0e10  &   20.9 	   & 	   3.3 	&   B93 	  \\ 
   108767  & $\delta$  CrV  &  295.5  &   46.1  &   26.9  &  B9.5V  &   3.4e10  &   --0.5 	   & 	   7.9 	&   W96 	  \\ 
  22468  &   HR   1099  &  184.9  &  --41.6  &   29.0  &  G5IV/K1IV  &   7.9e17  &   21.9 	   & 	  --2.3 	&   P97 	  \\ 
    &      &    &    &    &    &   1.6e17  &  8.2 	   & 	   --16.0 	&   P97 	  \\ 
    &      &    &    &    &    &   4.0e17  &   14.8 	   & 	  --9.4 	&   P97 	  \\ 
   161868  & $\gamma$  Oph  &   28.0  &   15.0  &   29.1  &  A0V  &   3.0e10  &  --33.1 	   & 	  --6.0 	& CCW97 	  \\ 
    &      &    &    &    &    &   4.2e10  &  --29.9 	   & 	  --2.8 	& CCW97 	  \\ 
   4128  &  $\beta$  Cet  &  111.3  &  --80.7  &   29.4  &  K0III  &   5.9e16  &   1. 	   & 	  --5.3 	&   P97 	  \\ 
    &      &    &    &    &    &  2.23e18  &   8. 	   & 	   1.7 	&   P97 	  \\ 
   120418  & $\theta$  Peg  &   67.4  &  --38.7  &   29.6  &   A2IV  &   3.0e10  &   --4.2 	   & 	   4.6 	& CLW98 	  \\ 
  358  & $\alpha$  And  &  111.7  &  --32.8  &   29.8  &  B8IVp  &  1.99e10  &   13.0 	   & 	   6.2 	& CLW98 	  \\ 
   8538  & $\delta$  Cas  &  127.2  &   --2.4  &   30.5  & A5III--IV  &  0.51e10  &   12.9 	   & 	   1.1 	&  LB92 	  \\ 
   120315  &   $\eta$  UMa  &  100.7  &   65.3  &   30.9  &  B3V  &   0.9e10  &   --3.4 	   & 	   2.1 	&  FW02 	  \\ 
 209952\tablenotemark{c}  & $\alpha$  Gru  &  350.0  &  --52.4  &   31.1  &   B7IV  &   2.0e10  &   13.0 	   & 	  --1.9 	&  CD95 	  \\ 
    &      &    &    &    &    &  1.12e10  &   10.2 	   & 	   0.9 	&  CD95 	  \\ 
   213558  & $\alpha$  Lac  &  101.3  &   --6.6  &   31.4  &  A1V  &   1.0e10  &  3.5 	   & 	   3.4 	&  LB92 	  \\ 
  88955  &   HR   4023  &  274.3  &   11.9  &   31.5  &  A2V  &  2.38e10  &   --1.7 	   & 	  --4.4 	& CLW98 	  \\ 
    &      &    &    &    &    &  3.98e10  &   10.5 	   & 	   7.8 	& CLW98 	  \\ 
   112413  &  $\alpha$2  CVn  &  118.3  &   78.8  &   33.8  &  A0spe  &  0.51e10  &   --1.9 	   & 	   2.2 	&   V93 	  \\ 
   177756  &  $\lambda$  Aql  &   30.3  &   --5.5  &   38.4  &   B9Vn  &  0.41e10  &  --21.9 	   & 	   3.7 	&  FW02 	  \\ 
    &      &    &    &    &    &  0.41e10  &  --26.5 	   & 	  --0.9 	&  FW02 	  \\ 
    &      &    &    &    &    &  0.96e10  &  --30.7 	   & 	  --5.1 	&  FW02 	  \\ 
   215789  & $\epsilon$  Gru  &  338.3  &  --56.5  &   39.7  &  A3V  &  1.58e10  &  --12.2 	   & 	  --4.3 	& CCW97 	  \\ 
    &      &    &    &    &    &  1.95e10  &   --7.2 	   & 	   0.7 	& CCW97 	  \\ 
    &      &    &    &    &    &  2.88e10  &   --1.1 	   & 	   6.8 	& CCW97 	  \\ 
   218045  & $\alpha$  Peg  &   88.3  &  --40.4  &   42.8  &  B9III  &   0.8e10  &   --4.7 	   & 	  --3.2 	&  FW02 	  \\ 
    &      &    &    &    &    &   1.1e10  &   --0.5 	   & 	   1.0 	&  FW02 	  \\ 
    &      &    &    &    &    &   0.6e10  &  2.5 	   & 	   4.0 	&  FW02 	  \\ 
 10144\tablenotemark{c,e}  & $\alpha$  Eri  &  290.8  &  --58.8  &   44.0  &  B3Vpe  &  0.50e10  &   18.9 	   & 	  16.2 	&  CD95 	  \\ 
    &      &    &    &    &    &  1.26e10  &   21.2 	   & 	  18.5 	&  CD95 	  \\ 
    &      &    &    &    &    &  1.86e10  &   11.0 	   & 	   8.3 	&  CD95 	  \\ 
    &      &    &    &    &    &  2.51e10  &  7.6 	   & 	   4.9 	&  CD95 	  \\ 
   141003  &  $\beta$  Ser  &   26.0  &   47.9  &   46.9  &   A2IV  &   1.0e10  &  --16.7 	   & 	   5.5 	&  FW02 	  \\ 
    &      &    &    &    &    &   2.9e10  &  --23.3 	   & 	  --1.1 	&  FW02 	  \\ 
    &      &    &    &    &    &   6.6e10  &  --20.7 	   & 	   1.5 	&  FW02 	  \\ 
   212061  & $\gamma$  Aqr  &   62.2  &  --45.8  &   48.4  &  A0V  &   7.94e9  &   --4.5 	   & 	   3.9 	& CCW97 	  \\ 
   135742  &  $\beta$  Lib  &  352.0  &   39.2  &   49.1  &  B8V  &   0.8e10  &  --33.7 	   & 	   --10.1 	&  FW02 	  \\ 
    &      &    &    &    &    &   6.4e10  &  --23.6 	   & 	   0.0 	&  FW02 	  \\ 
    &      &    &    &    &    &   6.4e10  &  --26.9 	   & 	  --3.3 	&  FW02 	  \\ 
   106625  & $\gamma$  CrV  &  291.0  &   44.5  &   50.0  &   B8IIIp  &   1.5e10  &  1.6 	   & 	   8.5 	&   W96 	  \\ 
    &      &    &    &    &    &   5.4e10  &   --2.0 	   & 	   4.9 	&   W96 	  \\ 
   148857  &  $\lambda$  Oph  &   17.1  &   31.8  &   50.9  &  A0Vpl  &   4.7e10  &  --24.8 	   & 	   1.5 	&  FW02 	  \\ 
   160613  &  $o$  Ser  &   13.3  &  9.2  &   51.5  &   A2Va  &   5.6e10  &  --29.0 	   & 	  --0.9 	&  FW02 	  \\ 
   222439  & $\kappa$  And  &  109.8  &  --16.7  &   52.0  &  B9IVn  &   1.1e10  &  0.8 	   & --4.2 	&   V93 	  \\ 
    &    &      &    &    &    &   6.1e10  &  7.6 	   & 2.6 	&   V93 	  \\ 
   186882  & $\delta$  Cyg  &   78.7  &   10.2  &   52.4  &   B9.5IV  &   2.8e10  &  --18.8 	   & 	  --6.9	&   W96 	  \\ 
     &      &    &    &    &    &   2.9e10  &  --16.3 	   & --4.4 	&   W96 	  \\ 
    &      &    &    &    &    &   3.2e10  &   --9.6 	   & 2.3 	&   W96 	  \\ 
  62044  & $\sigma$  Gem  &  191.2  &   23.3  &   55.5  &  K1III  &  5.85e17  &   32.0 	   & 	   8.9 	&   D97 	  \\ 
    &      &    &    &    &    &  7.50e17  &   21.4 	   & 	  --1.7 	&   D97 	  \\ 
   193924  & $\alpha$  Pav  &  340.9  &  --32.5  &   56.2  &   B2IV  &  1.24e10  &  --19.6 	   & 	  --2.8 	& CLW98 	  \\ 
    &      &    &    &    &    &  1.05e11  &  --18.6 	   & 	  --1.8 	& CLW98 	  \\ 
   213998  &   $\eta$  Aqr  &   66.8  &  --47.6  &   56.3  &  B9IV--Vn  &  2.70e10  &   --2.1 	   & 	   4.6 	&   V93 	  \\ 
   156928  &  $\nu$  Ser  &   10.6  &   13.5  &   59.3  &   A0/A1V  &   4.4e10  &  --27.7 	   & 	   0.4 	&  FW02 	  \\ 
 WD0501+527  &   G191--B2B  &   156.0   &   +7.1  &  68.8   &  DAw  &  1.86e18  &   19.3 	   & 	  --2.0 	&  Sa99 	  \\ 
    &      &    &    &    &    &  3.98e17  &  8.6 	   & 	   --12.7 	&  Sa99 	  \\ 
   111812  &   31  Com  &  114.9  &   89.6  &   90.9  &   G0II  &  7.66e17  &   --2.4 	   & 	   3.2 	&   D97 	  \\ 
  52089  & $\epsilon$  CMa  &  239.8  &  --11.3  &   132.  &  A1V  &   4.0e17  &  9.2 	   & 	   --10.2 	& GJ01\tablenotemark{f}	  \\ 
    &      &    &    &    &    &   6.0e17  &   16.2 	   & 	  --3.2 	&  GJ01 	  \\ 
\hline 
\enddata 
\tablenotetext{a}{Column Densities less than 10$^{13}$ are for \CaII,
and values greater than this are based on \NHI\ measurements.}
\tablenotetext{b}{References are listed in Table \ref{tab-ref}.}
\tablenotetext{c}{Star omitted from restricted sample used for determining \Vflow(93), see text.}
\tablenotetext{d}{$\alpha$ Eri is the brightest known Be emission line star, with evidence of variable chromosphere and variable stellar radial velocities \citep{PorriStalio:1988}.  } 
\tablenotetext{e}{Star with infrared excess indicating dust disk \citep[e.g.][]{Habingetal:2001}.  }
\tablenotetext{f}{Velocities are from GJ01 text and not GJ01 table.}
\end{deluxetable}

%% file: tab4.tex
\begin{deluxetable}{lllrll}
\tablecolumns{5} 
\tablewidth{0pc} 
\tablecaption{Regional Properties of Flow \label{tab-cld}}
\tablehead{ 
\multicolumn{2}{c}{Cloud\tablenotemark{a}}&\colhead{Vector}&\colhead{\dVi\ Range}&\colhead{Location\tablenotemark{b}}&\colhead{Stars Showing }\\
\multicolumn{2}{c}{Name}&\colhead{}&\colhead{(\kms)}&\colhead{$d_{\rm min}$,\glong,\glat}&\colhead{Cloud Component}\\
}
\startdata 
\hline 
1&LIC 	&	\Vflow(He)	&	0.0$\pm$1.5	&0 pc, \glong=150\deeg$\rightarrow$250\deeg\ &
All stars in interval \\
2\tablenotemark{c}&Blue Cloud	&	\Vflow(96)	&	--9.8$\pm$0.3	&$<$3 pc, 233\deeg$\pm$7\deeg,10\deeg$\pm$2\deeg &
$\alpha$ CMa, $\epsilon$ CMa \\
3&Aql/Oph	&	\Vflow(96)	&	--6.1$\pm$1.0	&$<$5 pc, 38\deeg$\pm$10\deeg,9\deeg$\pm$14\deeg&
$\alpha$ Aql, $\alpha$ Oph, $\zeta$ Aql, $\gamma$ Oph,  \\
&&&&&$\lambda$ Aql ($\delta$ Cyg) \\
4& Peg/Aqr &	\Vflow(96)	&	4.2$\pm$0.5	&$<$30 pc, 75\deeg$\pm$13\deeg,--44\deeg$\pm$5\deeg\ &
$\theta$ Peg, $\alpha$ Peg, $\gamma$ Aqr, $\eta$ Aqr\\
5&North Pole	&	\Vflow(96)	&	1.7$\pm$0.6	& $<$22 pc, \glat$>$+47\deeg, &
$\gamma$ UMa, $\alpha$ CrB, $\delta$ UMa, $\eta$ UMa,  \\
&&&&&$\alpha ^2$ CVn, $\beta$ Ser \\
6& South Pole: &	\Vflow(96)	&	2.4$\pm$1.0	&$<$3.6 pc, \glat$\leq$--48\deeg\ &
$\epsilon$ Ind, $\alpha$ Hyi, $\tau ^3$ Eri, $\beta$ Cet \\
7& Filament:	& \Vflow(96) & 6.7$\pm$1.8 & $<$22 pc, 284\deeg$\pm$12\deeg,--4\deeg$\pm$50\deeg\ & 
$\delta$ CrV, $\gamma$ CrV,  HR4023, $\delta$ Vel, \\
&&&&&$\alpha$ Hyi, $\epsilon$ Gru\\
\hline 
\enddata 
\tablenotetext{a}{Clouds marked with a ``:'' are poorly identified and uncertain.}
\tablenotetext{b}{$d_{\rm min}$ is the minimum cloud distance.}
\tablenotetext{c}{Cloud 2 is the ``Blue Cloud'' \citep{Lallementetal:1994,GryJenkins:2001}.}

\end{deluxetable} 

%% file: tab5.tex
\begin{deluxetable}{llllll}
\tablecolumns{6} 
\tablewidth{0pc} 
\tabletypesize{\small}
\tablecaption{LISM Heliocentric Velocity Vectors \label{tab-rev}} 
\tablehead{ 
\colhead{Vector} & \colhead{V} & \colhead{l} & \colhead{b} & \colhead{Basis} & \colhead{Ref.}  \\
\colhead{} & \colhead{\kms} & \colhead{\deeg} & \colhead{\deeg} & \colhead{} & \colhead{}  \\
}
\startdata 
Crutcher (C) &--28$\pm$2&25&+10&Optical \TiII, d$<$100 pc stars&\citet{Crutcher:1982} \\
Frisch \& York (FY) &--27&34&+15&Optical \CaII, d$<$100 pc stars& \citet{FrischYork:1986} \\
LIC & 26$\pm$1 & 6$\pm$3 & 16$\pm$3 & \CaII\ &  LB92, \citet{Bertinetal:1993} \\
G & 29 & 4.5 & 20.5 &  \CaII&  \citet{Bertinetal:1993} \\ 
Frisch (F) & --26.8$\pm$1.3 & 6.2 & 11.7 & \CaII, d$<$50 pc stars & \citet{Frisch:1995} \\
Bulk Flow (BF) & --28.1$\pm$4.6 & 12.4 & 11.6 & UV, Optical & This work \\
\enddata 
\end{deluxetable}

%% file: figurecaptions.tex
\clearpage

\begin{figure}[ht]
\epsscale{0.85}
\plotone{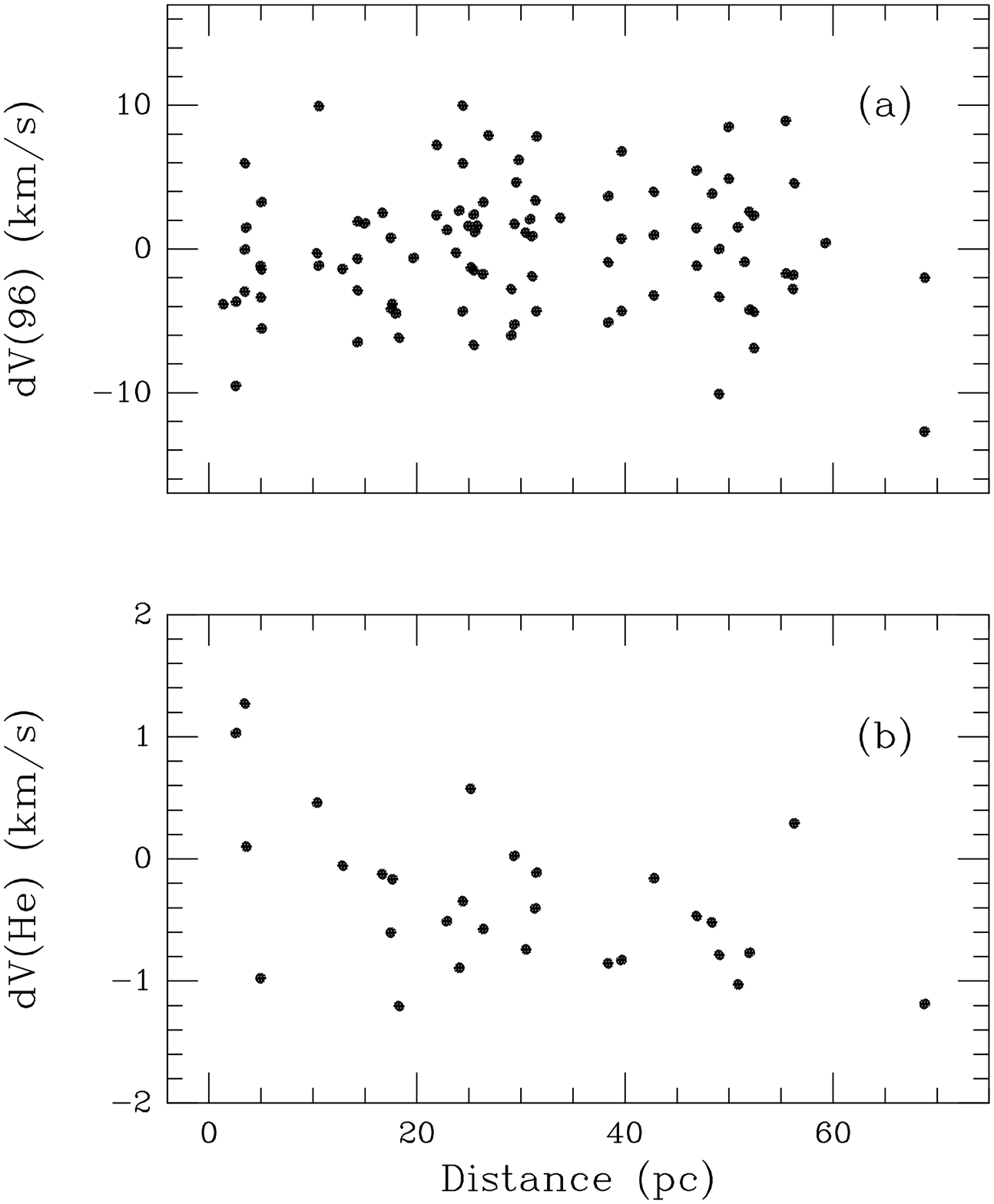}
\caption{
(a)  Dispersion (\dVi(96)) of component velocities in the rest frame of the CLIC
as defined by \Vflow(96) (see Table \ref{tab-vec}).
(b) Plot of \dVi(He) versus star distance, for components within 1.5 \kms\
of \Vflow(He) (i.e. $|$\dVi(He)$|  <$1.5 \kms).  }
\label{fig1}
\end{figure}

\begin{figure}[ht]
\plotone{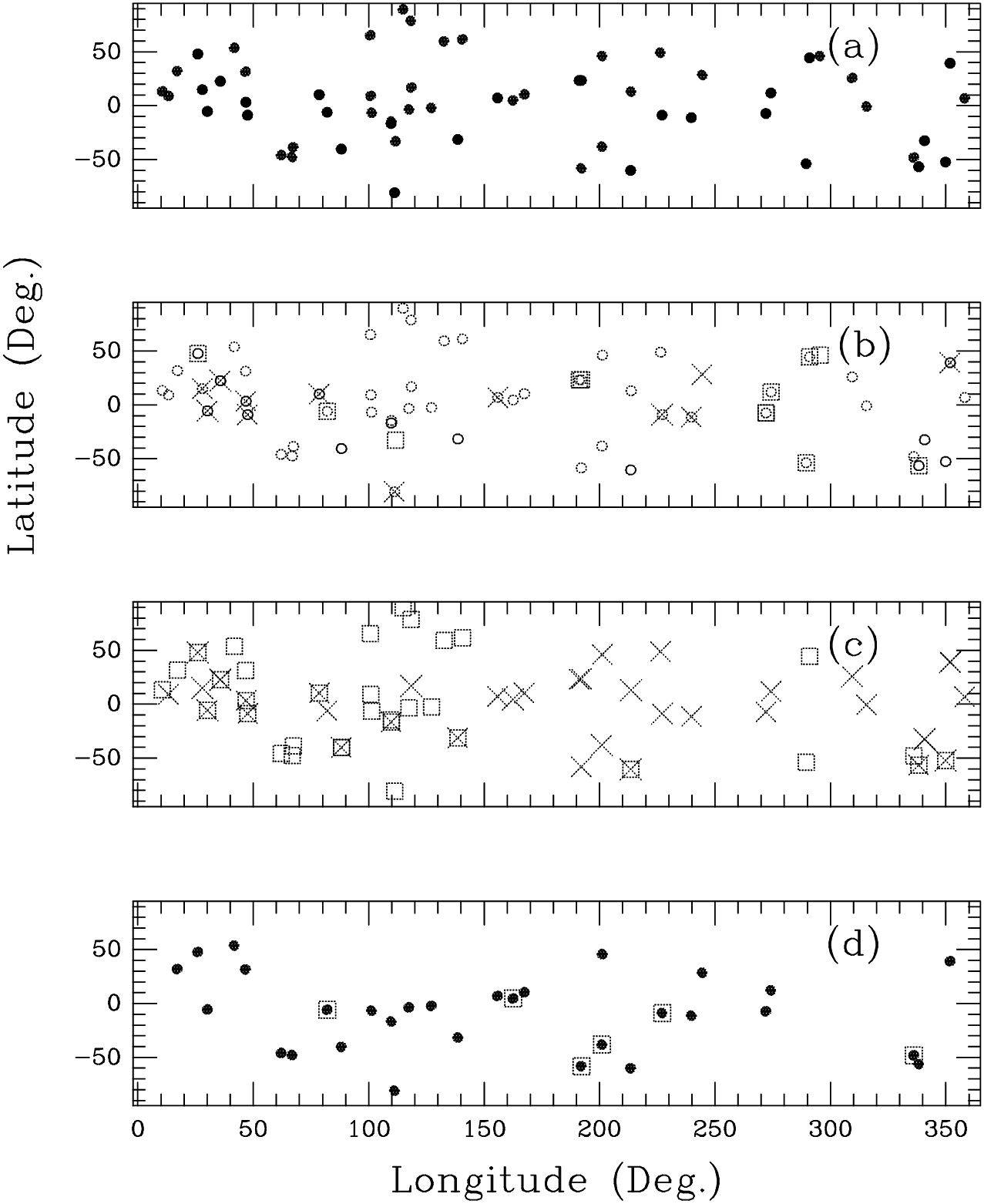}
\caption{
(a)  The galactic coordinates of the background stars used to derive \Vflow(96)
(Table \ref{tab-vec})1
(b)  Same stars as in (a), but coded for \dVi(96):  
circle -- $|\dVi(96)| <$ 5 \kms;
box -- $\dVi(96) >$ 5 \kms;
cross -- $\dVi(96) <$ --5 \kms.  Stars with multiple components show
more than one symbol.
(c)  The stars for which $|\dVi(96)| <$5 \kms\ are coded for 
the sign of \dVi(96):  box -- \dVi(96)$>$0 \kms; cross -- \dVi(96)$<$0 \kms. 
Stars with components in each range show both symbols.
(d)  The galactic coordinates of the 30 stars with components 
showing $|$\dVi(He)$|  <$1.5 \kms.  Boxes surround the positions of stars 
within 15 pc.}
\label{fig2}
\end{figure}

\begin{figure}[ht]
\plotone{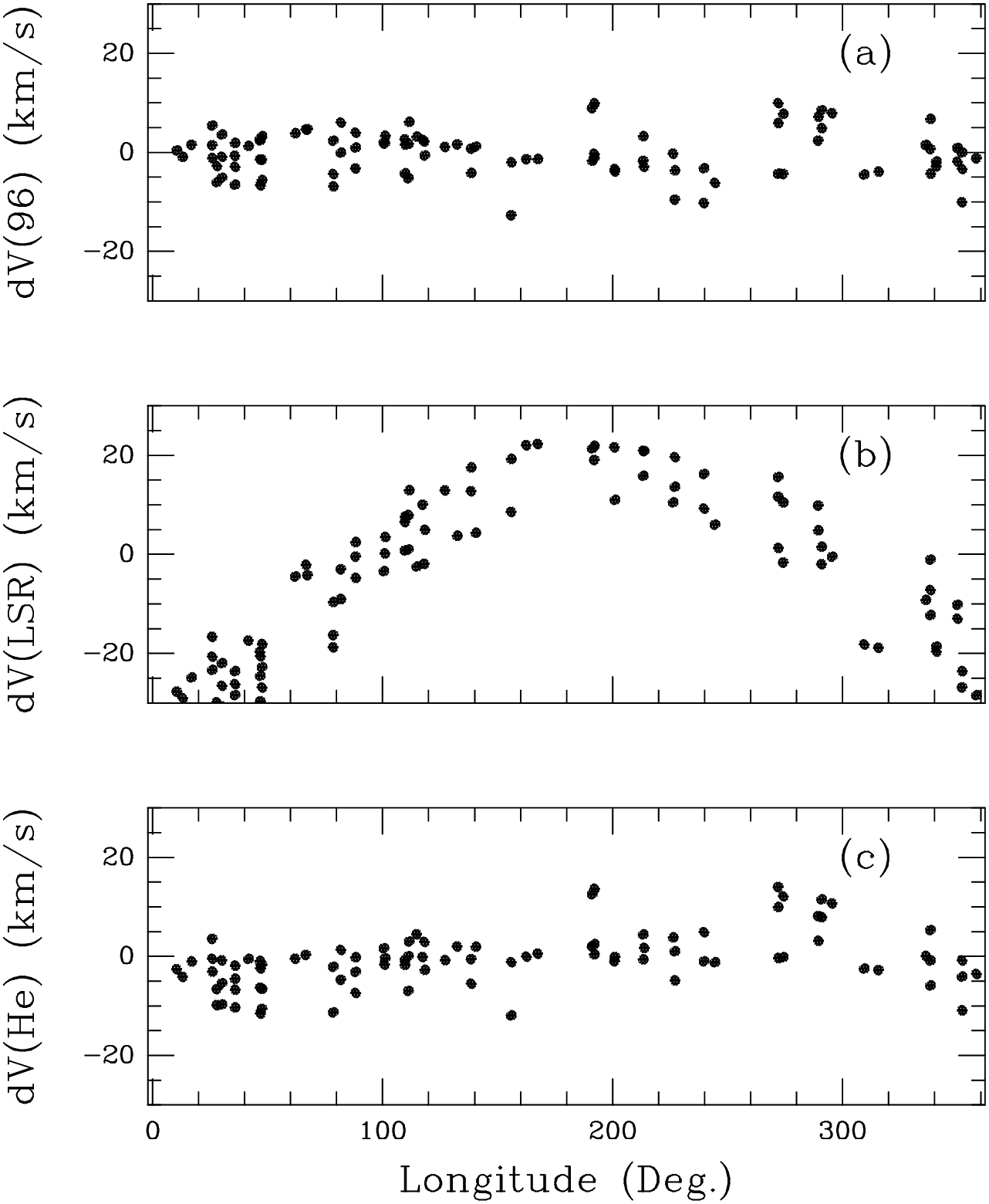}
\caption{
(a)  \dVi(96) plotted against galactic longitude for the restricted sample of stars.  
(b)  \dVi(LSR) plotted against galactic longitude for same stars.  The
sinusoidal appearance shows that the ISM within 30 pc of the Sun is not
near the rest velocity of the LSR (e.g. also see \citet{Frisch:1995}). 
(c)  \dVi(He) plotted against galactic longitude for same stars.   }
\label{fig3}
\end{figure}

\begin{figure}[ht]
\begin{center}
\includegraphics[angle=-90,width=5in]{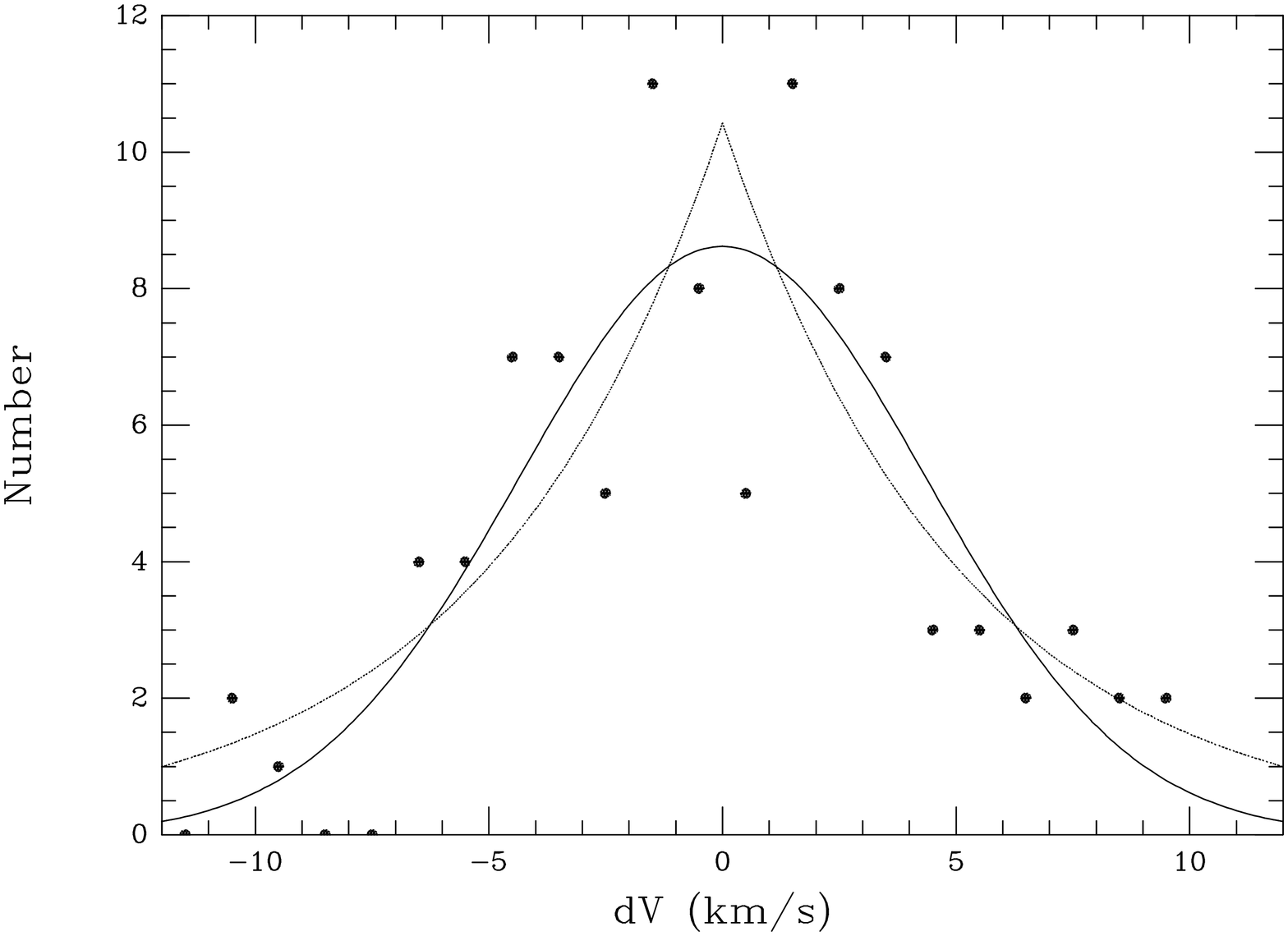}
\end{center}
\caption{
Histogram of \dV\ values, calculated used \Vflow(96).  
A best-fit exponential
component distribution (dotted line) is overplotted for $\eta$=5.1 (Equation \ref{eq-psi}).
The best-fit Gaussian distribution (solid line) for these same components, 
with $b$=6.2 \kms, is also shown (Equation \ref{eq-gaus}).}
\label{fig4}
\end{figure}

\begin{figure}[ht]
\epsscale{1.0}
\caption{
The \Vflow(96) upstream direction in the LSR (\glong=1.8\deeg, \glat=--3.2\deeg) 
is shown plotted on a map of 21 cm emission integrated over the velocity 
range --450 to 400 \kms, with the \HI\ Loop I faintly visible \citep{Hartmann:1997}. 
The Loop I position in 408 MHz radio continuum emission is 
overplotted as two lines \citep[from][]{Berkhuijsen:1971,Haslam:1982}. 
The LSR upstream direction of \Vflow(96) (dot) is 
$\sim$20\deeg\ from the tangential direction of Loop I, consistent with a
CLIC origin associated with either an expanding Loop I superbubble, or an
outflow from the Sco-Cen Association \citep[e.g.][]{Frisch:1995}. }
\label{fig-sxrb}
\end{figure}

\begin{figure}[ht]
\epsscale{1.0}
\includegraphics[angle=-90,width=7.5in]{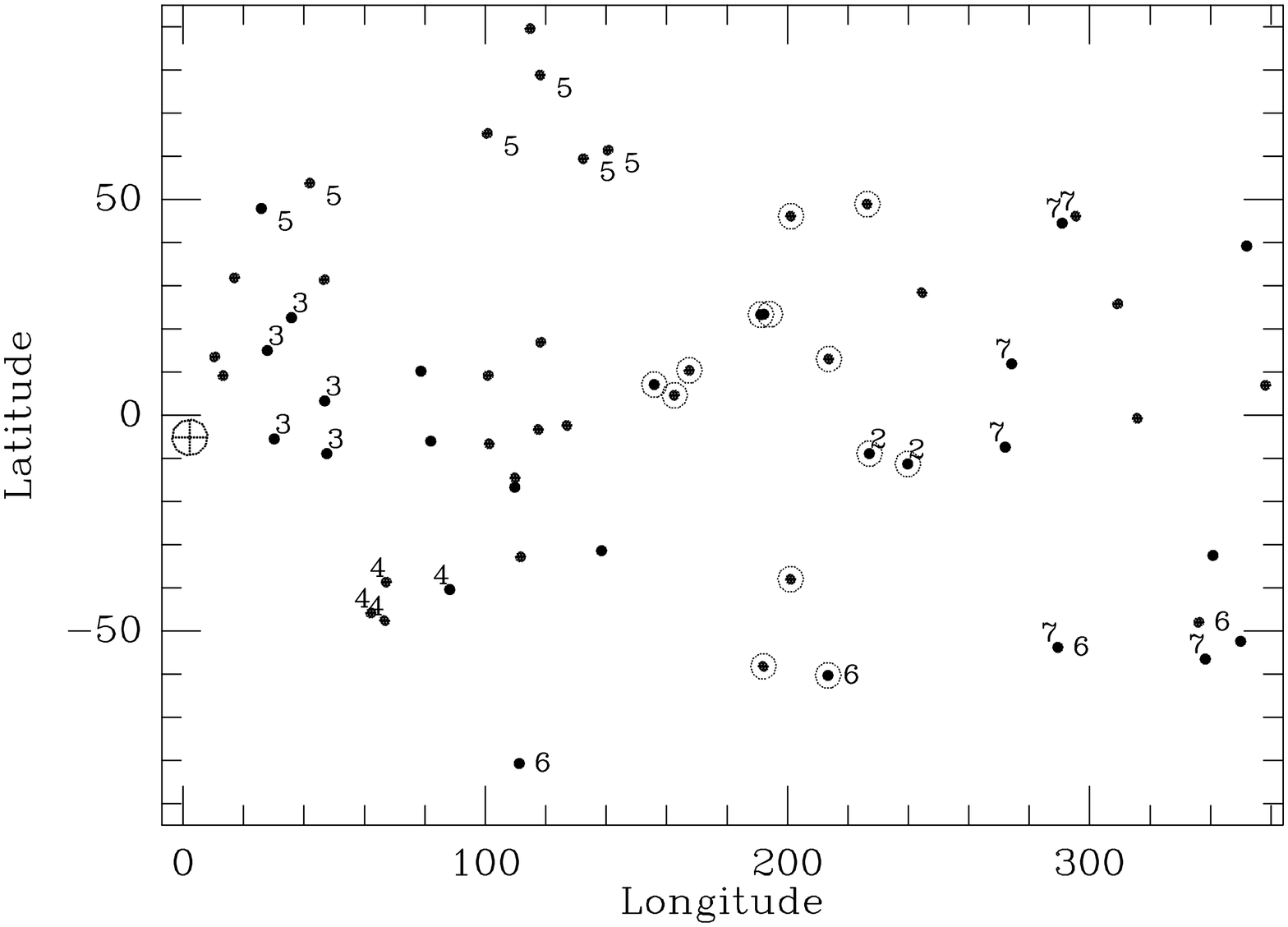}
\caption{The Clouds listed in Table \ref{tab-cld} are shown plotted
against positions of stars in the restricted sample.  
Positions of stars showing Cloud 1 components are circled, and other
clouds are identified by cloud number.
The LSR upstream direction of \Vflow(96) is identified by the circled cross.
}
\label{fig6}
\end{figure}

\begin{figure}[ht]
\epsscale{1.0}
\plotone{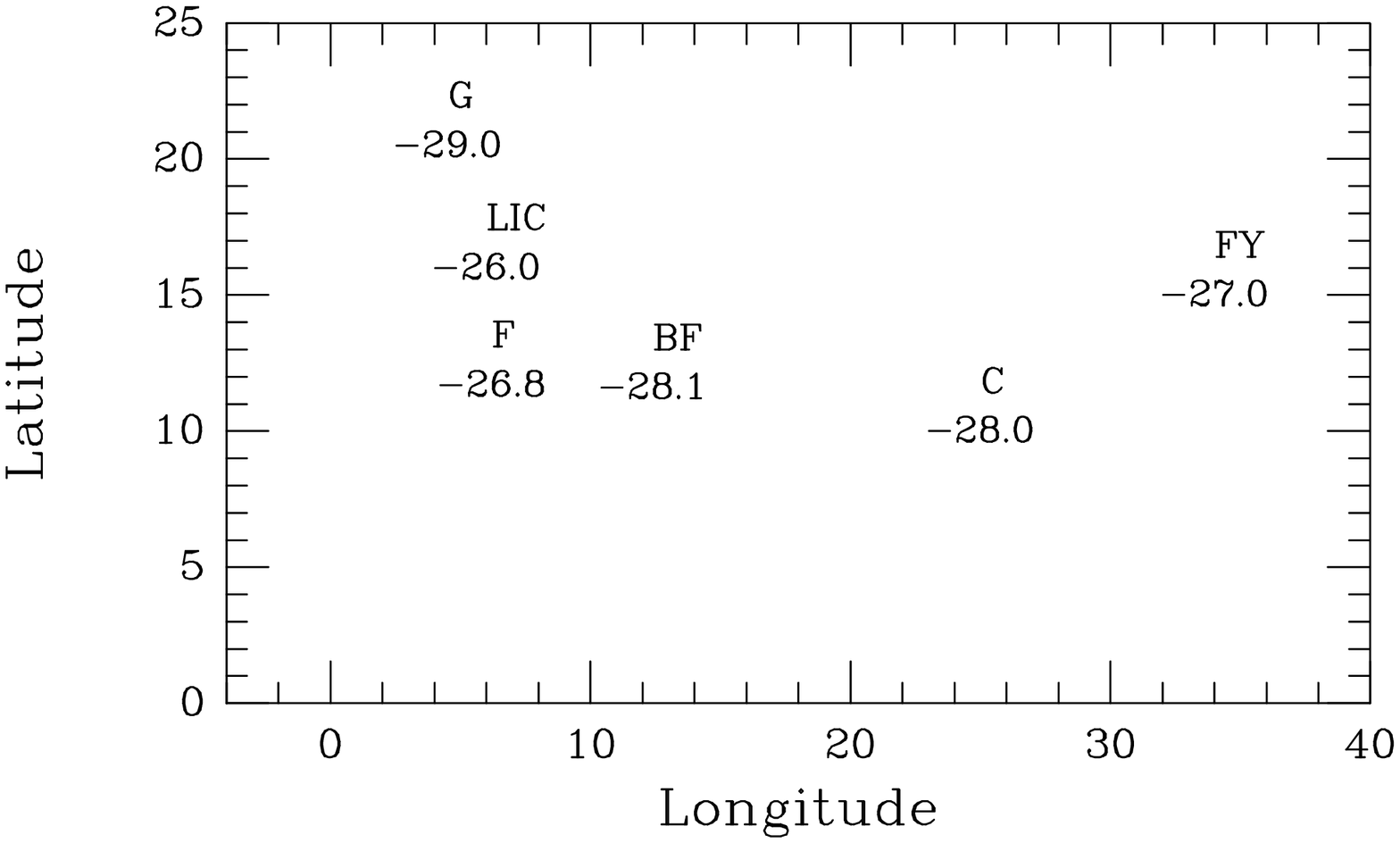}
\caption{Heliocentric upstream directions for various
bulk velocity vectors found for nearby ISM
are plotted.  Labels and data for individual vectors are presented in Table \ref{tab-rev}.  }
\label{figcom}
\end{figure}